\documentclass[%
 reprint,
 amsmath,amssymb,
 aps,
 prmaterials,
 groupedaddress,
]{revtex4-2}

\usepackage{graphicx}% Include figure files
\usepackage{dcolumn}% Align table columns on decimal point
\usepackage{bm}% bold math

\usepackage{comment} 
\usepackage{booktabs}
\usepackage{xcolor} 
\begin{document}
\preprint{APS/123-QED}

%%% Title
\title{Predicting THz Generation Capability of Organic Crystals through Data Mining and Crystal Nonlinearity Models}

%%% Author List
\author{Enoch Sin Hang Ho}
\email{enochho@byu.edu}
\affiliation{%
 Department of Chemistry and Biochemistry, Brigham Young University\\
}%
\author{Ashton Roma}%
\author{Connor Barlow}
\author{Matthew Lutz}
\author{Natalie Green}
\author{Stacey Smith}
\author{David Michaelis}
\email{dmichaelis@chem.byu.edu}
\author{Jeremy A Johnson}%
\email{jjohnson@chem.byu.edu}
\affiliation{%
 Department of Chemistry and Biochemistry, Brigham Young University\\
}%

%%% Abstract
\begin{abstract}
We report the use of DFT computation and mathematical models to predict the nonlinear dielectric polarization ($P^{NL}$) and nonlinear susceptibility coefficients ($\chi^{(2)}_{IJK}$) of organic materials based on their crystal structures. We apply this computation approach on single-component crystals, co-crystals and ionic crystals found through data mining the Cambridge Structural Database. We verify these computational results with experimental terahertz (THz) generation efficiencies for known THz generators, demonstrating consistency between the measurements and the computed values.  Several mined structures show similar or larger $P^{NL}$ values compared to state-of-the-art THz generation crystals DAST, OH1 and BNA, suggesting great potential for their use in nonlinear optical (NLO) applications. Importantly, we also compared the resulting model of nonlinear optical tensor components with commonly-used simplifications of nonlinearity, showing that the comprehensive approach should be the standard method to evaluate the nonlinear optical properties of single-crystalline materials.

\end{abstract}

\maketitle

%%% Main Text
\section{Introduction}
Solid-state and single crystalline materials are essential to many fields due to their distinctive and diverse properties enabled by their specific atomic arrangement. These fields include catalysis \cite{Li_2020}, biomaterials \cite{Brown_2023}, energy storage \cite{Ye_2020}, and electronics and semiconductors \cite{Eswaran_2025}, which can benefit from solid-state properties, including electrical conductivity, magnetism, thermal expansion and chemical properties. 

One important application of solid-state materials is their use as nonlinear optical materials for  frequency conversion of laser light \cite{Franken_1961}, such as for the generation of high intensity broadband terahertz (THz) frequency light. In this work, we demonstrate the use of DFT computation in tandem with crystal models of optical nonlinearity to predict the efficiency of frequency conversion in crystalline materials. In particular, we show that computing the nonlinear susceptibility coefficients ($\chi^{(2)}$) and the nonlinear dielectric polarization ($P^{NL}$) of organic materials allows us to predict their potential for THz generation based on their reported crystal structures.

When a nonlinear optical (NLO) material is irradiated with coherent laser light, the NLO response of the material produces an altered electric field with new frequencies not present in the original radiation. This optical nonlinearity is closely related to the structure and atomic makeup of materials---the molecular building blocks in organic materials. One key molecular property is the molecular hyperpolarizability ($\beta_{ijk}$), which is the first nonlinear term of the series expansion of the dipole moment ($\mu$) shown in Eq. \ref{eq:2mu} \cite{Goncharov_2012, Verbiest_1997}.
\begin{equation}
    \mu_i = \mu_{i0} + \sum_{j}\alpha_{ij} E_j + \sum_{jk}\beta_{ijk} E_j E_k + ...
    \label{eq:2mu}
\end{equation}

This equation shows how an external electric field ($E_{i,j,k}$) can induce a change in the static molecular dipole moment ($\mu_{i0}$) through the linear response from the molecular polarizability ($\alpha_{ij}$) and nonlinear response from the molecular hyperpolarizability ($\beta_{ijk}$). The larger the $\beta_{ijk}$-value of the molecular building blocks comprising the crystal, the stronger the molecular dipole moment responds nonlinearly to an incident electric field. 

As key molecular properties like  $\beta_{ijk}$ are directional in nature, molecular alignment due to crystal packing is also a crucial factor. To manifest a second-order optical nonlinearity, an NLO crystal must be non-centrosymmetric \cite{Ivanova_2010, Boyd_2008}. In centrosymmetric structures, any inherent nonlinearity induced in one molecular component is negated by inversion symmetric counterparts. Similar to the molecular dipole nonlinearity, a series expansion is also applied to crystal polarization density ($P_I$) as shown in Eq. \ref{Ch3:PFull} \cite{Goncharov_2012, Verbiest_1997}.

\begin{equation}
    P_{I} = \varepsilon_{0}(\sum_{J}\chi_{IJ}^{(1)}E_J + \sum_{JK}\chi_{IJK}^{(2)}E_JE_K + ...)
    \label{Ch3:PFull}
\end{equation}

$P_{I}$ can be modulated linearly through the 1$^{st}$-order electric susceptibility $\chi^{(1)}_{IJ}$, and nonlinearly through the 2$^{nd}$-order electric susceptibility $\chi^{(2)}_{IJK}$ and higher order terms. 
$\chi^{(2)}_{IJK}$ is a tensor that quantifies the strength of the nonlinear response to a macroscopic electric field. Thus, it is natural to model $\chi^{(2)}_{IJK}$ using the microscopic molecular hyperpolarizability together with the bulk crystal packing. 
The $\chi^{(2)}_{IJK}$ tensor can be approximated using Eqs. \ref{Ch3eq:d_Full} and \ref{Ch3:chi2_d} as follows \cite{PhysRevA.26.2028}:
\begin{multline}
d_{IJK}(0\omega;\omega,\omega) = Nf_I^{0\omega}f_J^{\omega}f_K^{\omega} \\
\times \sum_{\substack{s=1 \\ i,j,k}}^{n} cos(I,i(s))cos(J,j(s))cos(K,k(s)) \beta_{ijk} 
    \label{Ch3eq:d_Full}
\end{multline}
\begin{equation}
    \chi^{(2)}_{IJK} = 2d_{IJK}
    \label{Ch3:chi2_d}
\end{equation}

The $d_{IJK}$ tensor is a traditional/historic tensor that relates to the $\chi^{(2)}_{IJK}$ tensor by a factor of 2 and is still commonly used to describe the nonlinearity of crystal structures. The cosine and $\beta$ terms in the summation of Eq. \ref{Ch3eq:d_Full} represent the sum of all projections of hyperpolarizability vectors of each molecular building block ($n$) from molecular coordinate ($i$, $j$, $k$) onto the crystallographic coordinate ($I$, $J$, $K$). $N$ is the density of the unit cell, and the $f$ terms are the local field factors, related to the refractive index of the material. In centrosymmetric materials, the overall $\chi^{(2)}_{IJK}$ is zero because of the complete cancellation of the molecular hyperpolarizability vectors due to the inversion symmetry. 

In contrast, non-centrosymmetric structures lack a center of inversion; therefore the complete cancellation of the $\beta_{ijk}$ vectors is not structurally mandated, permitting nonzero $\chi^{(2)}$ values. (Certain non-centrosymmetric crystals can still exhibit complete cancellation of $\beta_{ijk}$ vectors due to their crystallographic symmetry.) The strength of $\chi^{(2)}_{IJK}$ therefore depends on the density of chromophores, their molecular hyperpolarizabilities, and the crystal packing. The highest possible value of $\chi^{(2)}_{IJK}$ is achieved when the molecular building blocks are packed in a linear head-to-tail geometry. This is because the $\beta_{ijk}$ vectors will be summed as a simple addition in Eq. \ref{Ch3:chi2_d}, rather than a sum of the projection of the $\beta_{ijk}$ vectors. The value of $\chi^{(2)}_{IJK}$ will be less than the highest possible value for any other forms of packing that diverge from this ideal head-to-tail geometry. Because of this directional requirement for crystal packing, the nonlinear optical response of an NLO crystal also depends critically on the polarization of the incident light and the specific crystallographic direction it interacts with \cite{hubner1994, dmitriev1991}. When including directionality in Eq. \ref{Ch3:PFull}, and only retaining the nonlinear term, the induced nonlinear crystal polarization can be written as:
\begin{gather}
 \begin{bmatrix}
 P_1 \\
 P_2 \\
 P_3 \\
 \end{bmatrix}^{NL}
 =
 2\epsilon_{0}
  \begin{bmatrix}
  d_{11} & d_{12} & d_{13} & d_{14} & d_{15} & d_{16} \\
  d_{21} & d_{22} & d_{23} & d_{24} & d_{25} & d_{26} \\
  d_{31} & d_{32} & d_{33} & d_{34} & d_{35} & d_{36} \\
   \end{bmatrix}
  \begin{bmatrix}
  E_1^{2} \\
  E_2^{2} \\
  E_3^{2} \\
  2E_2E_3 \\
  2E_1E_3 \\
  2E_1E_2 \\
   \end{bmatrix}
   \label{Ch3_5P_vector}
\end{gather}

The $d_{IL}$ terms are the $d_{IJK}$ tensor in contracted notation, and the correspondence between the indices are:

\begin{center}
\begin{tabular}{ c c c c c c c}
 $JK$: & ~11~ & 22~ & 33~ & 23,32 & 13,31 & 12,21 \\ 
 $L$: & ~1~ & 2~ & 3~ & 4 & 5 & 6
\end{tabular}
\end{center}

The $P_i^{NL}$ terms in Eq. \ref{Ch3_5P_vector} are the induced nonlinear crystal polarizations in the three defined directions. Through proper projections, the $P_i^{NL}$ terms can be computed in the crystallographic coordinate or in Cartesian coordinate. (The coordinate transformation between the two coordinate systems is shown in Eq. \ref{Ch3eq:Fractional2Cartesian} and \ref{Ch3eqf2c} in the Supplementary Information.) As previously discussed, the nonlinear optical responses are anisotropic across different crystallographic directions. Therefore, with proper consideration of the directions of the incident electric field and the appropriate elements of the $\chi^{(2)}_{IJK}$ (or the $d_{IJK}$ tensor), the total induced nonlinear polarization of a crystalline material for light incident on a specific crystal face can be computed as a vector sum of the relevant $P_i^{NL}$ terms. This induced polarization value serves as a key indicator to evaluate the strength of the nonlinear optical response of an NLO crystal.

The $\chi^{(2)}_{IJK}$ elements can be measured with a few methods, such as the Maker fringe technique \cite{Jerphagnon_1970}. However, such measurements can only be performed on well-developed crystals where crystal growth in different crystallographic directions is possible \cite{Notake2019}. At least 3 different crystals grown in three different crystallographic directions are required in order to obtain all the tensor elements. Therefore, it is not always practical to obtain all the  $\chi^{(2)}$ values through such measurements even for already-existing NLO crystals. In contrast, from a material development perspective, the ability to evaluate the theoretical performance of materials prior to development is crucial. This strategy significantly saves time and resources by directing synthesis, crystal growth, and processing resources only to high-potential candidates.

\begin{figure}[h]
    \centering
    \includegraphics[width=1\linewidth]{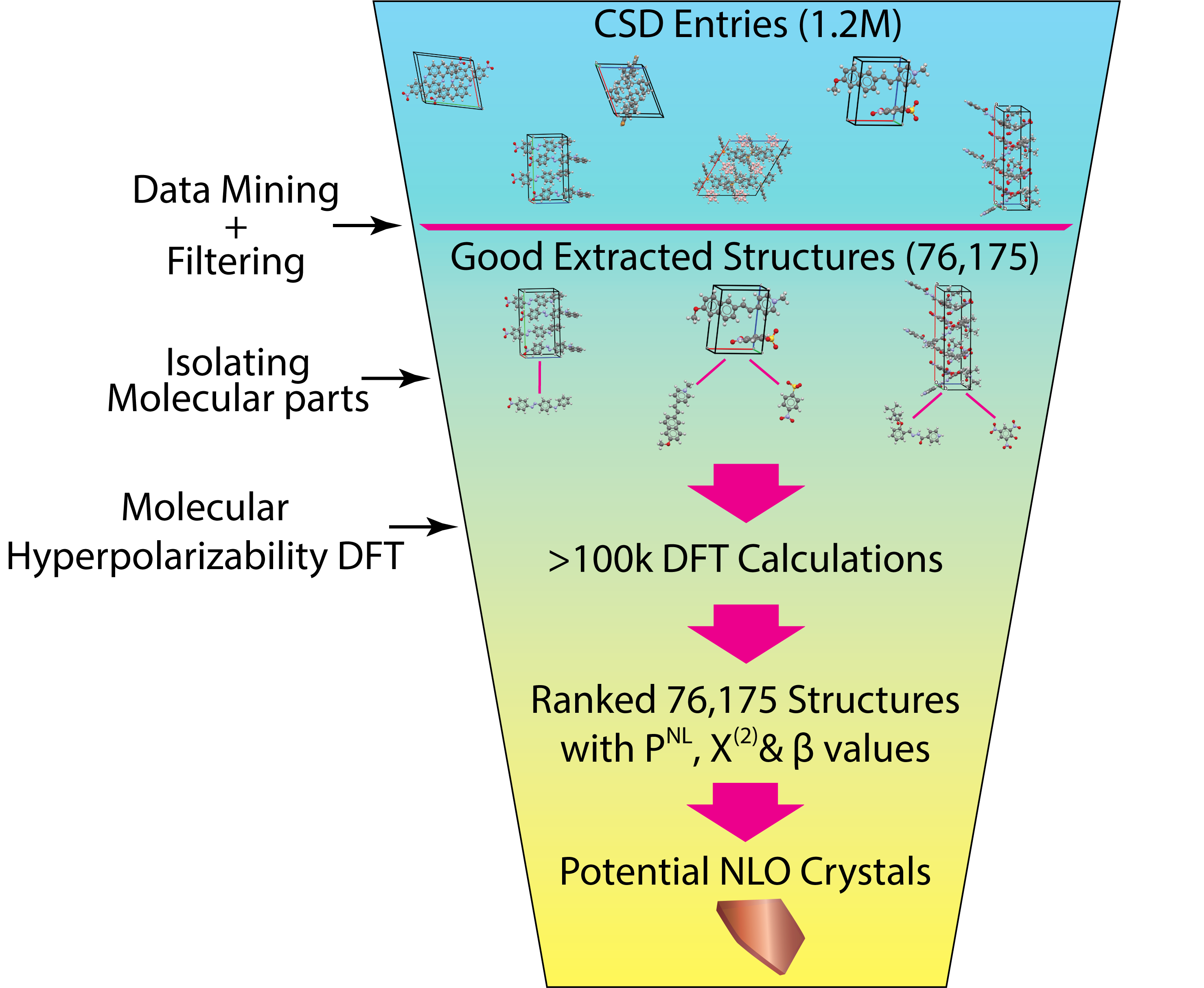}
    \caption{Workflow of the combined data mining/computational process to predict the nonlinearity of organic crystalline structures for intense THz generation.}
    \label{fig:DM_Scheme}
\end{figure}

To enhance our ability to estimate the NLO performance of known materials, we significantly expanded and improved our previous framework \cite{Valdivia-Berroeta_2022} on the identification of new high-performing NLO materials (See Fig \ref{fig:DM_Scheme}). We have expanded our dataset of X-ray determined crystal structures extracted from the Cambridge Structural Database (CSD) from 14,609 to 76,175. In addition to a nearly fivefold larger dataset, we now calculate all the $\chi^{(2)}_{IJK}$ tensor elements and the induced crystal nonlinear polarization. The calculated crystal nonlinear polarization serves as the indicator to predict the NLO performance of crystals, enabling the identification and ranking of  promising NLO candidates. To verify our computational framework, we compared the calculated NLO properties of known NLO crystals with experimental measurements in the context of THz generation via optical rectification---a rapidly expanding area of nonlinear optics \cite{EnochHo2025, Yang_2023,XUAN2025141180}.

%%%%%%%%%%%%%%%%%%%%%%%%%%%%%%%%%%%%%%%%%%%%%%%%%%%%%%%%%%
%%%%%%%%%%%%%%%%%%%%%%%%%%%%%%%%%%%%%%%%%%%%%%%%%%%%%%%%%%
\section{Experiment}

\subsection{Data Mining}
From over 1.2 million structures in the CSD, we extracted 77,843 non-centrosymmetric crystals as candidates for NLO application---a dataset significantly larger than 14,609 candidates extracted in our previous work \cite{Valdivia-Berroeta_2022}. Previously, we mined non-centrosymmetric structures by applying strict structural filters. Specifically, we restricted our selection to single-component crystals composed of only neutral molecules, excluding ionic and co-crystalline structures due to the ease of synthesis and processing. We also constrained the molecular building blocks to those containing only H, C, N, S, O, P, F, Cl, or Br atoms. To expand our dataset, we have now included ionic structures and co-crystals while removing the atomic type constraints. Ionic structures were specifically incorporated because some well-known THz generators are ionic, including DAST, DSTMS, and HMQ-TMS \cite{Vicario:15, Lu:15}. During this updated CSD data mining process, both crystallographic and molecular information were retrieved for all structures matching the expanded criteria.

%%%%%%%%%%%%%%%%%%%%%%%%%%%%%%%%%%%%

\subsection{2\textsuperscript{nd} Order Nonlinear Coefficients}
The $\chi^{(2)}$ tensors for all extracted structures were calculated with Eqs. \ref{Ch3eq:d_Full}-\ref{Ch3:chi2_d}, using the molecular hyperpolarizability tensors computed through DFT calculation and extracted crystallographic information, such as space group symmetry, unit cell dimensions, angles and volumes. For ionic and co-crystalline structures, the extracted molecular entries are comprised of multiple components (ions or co-formers) rather than a single molecular entity. In these cases, we needed to perform DFT calculations on each individual molecular entity. Including multiple components in a single molecular DFT calculation often triggers spurious inter-component charge transfer, which can artificially inflate the calculated values by orders of magnitude. This artifact occurs when non-bonded molecules are incorrectly treated as a single system \cite{Peach_2008,Sekino_2007,Yanai_2004}. To preclude this issue, we partitioned the constituent units (individual ions for ionic systems and individual chromophores for co-crystalline structures) and performed independent DFT calculations to compute the molecular hyperpolarizability tensors. Additionally, the local field factors in Eq. \ref{Ch3eq:d_Full} for all the extracted structures were approximated using the refractive indices of DAST instead of the crystal-specific data. Since the refractive index values are unavailable for the vast majority of the extracted structures, we applied the refractive index of DAST---a well-established NLO crystal---as the sole uniform constant in these calculations to ensure comparability. Details of DFT calculations and the subsequent $\chi^{(2)}_{IJK}$ tensors calculation are provided in the supplementary information.
%%%%%%%%%%%%%%%%%%%%%%%%%%%%%%%%%

\subsection{Nonlinear Optical Crystal Polarization}
The final step is to calculate the induced nonlinear polarization using Eq. \ref{Ch3_5P_vector}. The two necessary inputs are the $\chi^{(2)}_{IJK}$ tensors (obtained from the previous step) and the electric field input of the incident light source. Different crystal orientations exhibit significant variations in the nonlinear optical response due to the differences in relative orientation between the molecular alignment and incident light. Therefore, we systematically examine various combinations of the polarization of the irradiation source and the three primary crystal faces. For example, for an x-cut crystal, the input light propagates in the x-direction regardless of the polarization of the input light, thus the crystal polarization induced in the x-direction ($P_x$) has no effect on the oscillating electric field of the input light. This means that only the crystal polarizations induced in the y- and z-direction ($P_y$, $P_z$) contribute to the overall induced crystal polarization, thus $P_{x-cut}^{NL}= \sqrt{P_y^2+P_z^2}$. Furthermore, the values of $P_y$ and $P_z$ were calculated by considering various polarized light input, namely y-polarized, z-polarized, or yz-polarized input light.  Three induced crystal polarizations were calculated for each crystal face, resulting in a total of 9 different induced polarizations from all three main crystal faces.  Lastly, the maximum value of the induced polarizations ($P_{max}$) serves as the indicator to quantify the crystals' nonlinearity for ranking purposes. The crystal face corresponding to this maximum value was also recorded, providing a guide for the growth orientation required to achieve the optimal nonlinear response. 
%%%%%%%%%%%%%%%%%%%%%%%%%%%%%%%%%%%%%%%%%%%%%%%%%%%%%%%%%%
%%%%%%%%%%%%%%%%%%%%%%%%%%%%%%%%%%%%%%%%%%%%%%%%%%%%%%%%%%

\section{Results and Discussion}

\subsection{Data Mining}
\begin{figure}[h]
    \centering \includegraphics[width=1\linewidth, trim={0 0 0 0},clip]{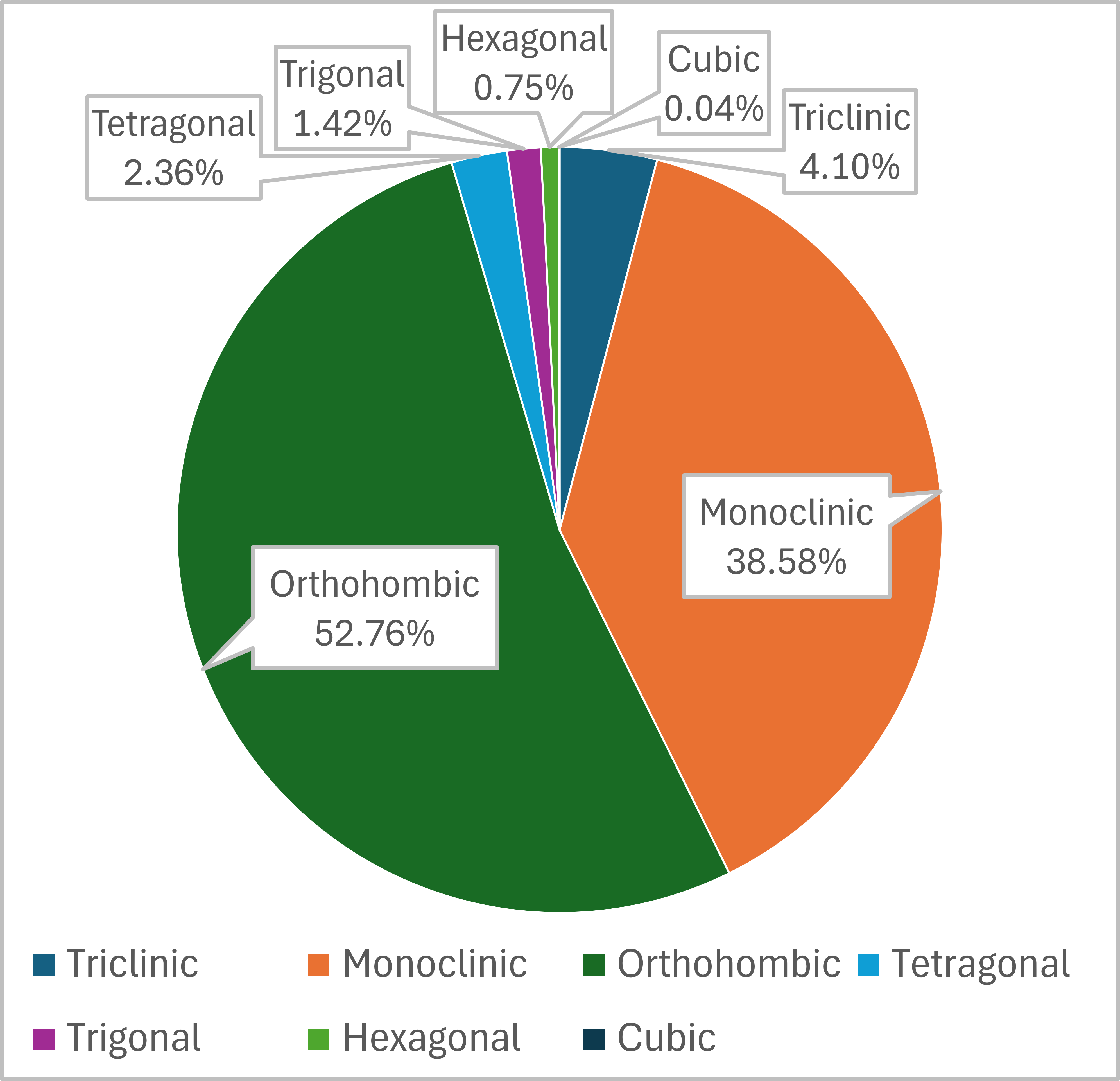}
    \caption{Crystal system statistics, showing most of the non-centrosymmetric crystals mined from the CSD are in the lower-symmetry crystal systems (monoclinic and orthorhombic). }
    \label{DM_stat_pie_chart}
\end{figure}

From 1.2 million structures in the CSD 2023.1 database, 77,843 structures were extracted after applying the selective criteria. Some structures contain positional disorders, resulting in failure of the DFT calculations. For structures with minor disorder, such as partially occupied atoms, we modified the structural files to retain only the main component of the disorder. Structures with significant disorder---such as atomic positions overlapping and unreasonable interatomic distances---were excluded. This processing yielded a final dataset of 76,175 candidate structures. Among the remaining extracted structures, 23,530 structures are composed of a single type of neutral molecule, and 52,645 structures are ionic or co-crystalline. Among all seven crystal systems, the two crystal systems contributing the largest fractions of the non-centrosymmetric structures are orthorhombic (52.76$\%$) and monoclinic (38.58$\%$) as indicated in Figure \ref{DM_stat_pie_chart}. This result aligns with expectations, as organic crystals typically pack into lower-symmetry space groups due to their larger molecular volume compared to inorganic crystals.

%%%%%%%%%%%%%%%%%%%%%%%%%%%%%%%%%%%%%%%%%%%%%%%%%%%%%%%%%%

\subsection{Induced Polarization and Nonlinear Coefficients Calculation}
Table \ref{Ch3:Cal_d} summarizes the calculated properties and crystallographic information for various known THz generators, THz generators identified through our previous data mining effort \cite{Valdivia-Berroeta_2022},  and newly identified potential NLO crystals. The calculated properties and crystallographic information include the induced dielectric polarization density ($P_{max}$), corresponding crystal face and the nonlinear optical coefficients ($d_{IJK}$). To verify the accuracy of our model predictions, 1) we compare our calculated $d_{IJK}$ values to experimentally measured values for DAST and OH1 and the magnitude to the inorganic crystals, 2) we examine the tensor structure of the computed $d_{IJK}$ elements, and 3) we compare the magnitudes of the $P{max}$ predictions with measured THz output efficiencies.

First, we compared our calculated $d_{IJK}$ values to reported experimental values. The calculated $d_{111}$ and $d_{122}$ for DAST are 309 pm/V and 47 pm/V, showing excellent agreement with the experimental values (290$\pm$15 pm/V and 41$\pm$3 pm/V respectively at 1542 nm) \cite{Meier_1998}. Similarly, the calculated $d_{333}$ (160 pm/V) and $d_{322}$ (18 pm/V) for OH1 are in great agreement with reported $d_{333}$ and $d_{322}$ for OH1 (120$\pm$10 pm/V and 13$\pm$2 pm/V at 1900 nm) \cite{Li_2014, Hunziker:08}. Furthermore, comparing our calculated values for organic crystals to the reported values for state-of-the-art inorganic NLO crystals, such as BBO ($d_{22}$ = 2.2 pm/V at 1064 nm) \cite{KLEIN_2003} and LiNbO\textsubscript{3} ($d_{32}$ = 25.2 pm/V and $d_{31}$ = 4.6 pm/V at 1064 nm) \cite{Shoji_97}, the calculated values for the organic nonlinear optical crystals are all larger. These results align with the observation that organic NLO crystals generally have higher nonlinear susceptibilities than these inorganic crystals. 

\begin{table*}
\caption{Calculated maximum induced dielectric polarizations ($P_{max}$), corresponding crystal faces and generation axis, space groups, and nonlinear optic coefficients ($d_{IJK}$) of known nonlinear optical crystals (DAST, OH1, MNA), previous identified crystals through our work (PNPA, NMBA and ZPAN), and selected newly identified crystals.}
\label{Ch3:Cal_d}
\begin{ruledtabular}
\begin{tabular}{c c c c c c c c}
ID & Type & $P_{max}$ & Corresponding   & Generation  & Space  & \multicolumn{2}{c}{Calculated  $d_{ijk}$} \\
   & &  (mC/m$^{2}$)&  Crystal Face &  Axis &  Group & \multicolumn{2}{c}{(pm/V)}  \\ \hline
\multicolumn{8}{c}{Known THz Generators} \\
DAST& Ionic & 54.9 & (010), (001)& [100] & $Cc$ & $d_{111}$ = 309  & $d_{122}$ = 47  \\
& & & &  &  & (290±15)\footnotemark[1]& (41±3)\footnotemark[1]\\
OH1& Neutral & 28.4 & (010), (100)&[001]& $Pna2_1$ & $d_{333}$ = 160 & $d_{223}$ = 18 \\
 & & & & & & (120±10)\footnotemark[2] &(13±2)\footnotemark[2]\\
MNA& Neutral &12.5& (010)&[101]& $Cc$ & $d_{111}$ = 66& $d_{133}$ = 7\\
\bottomrule
\multicolumn{8}{c}{THz Generators identified through our previous data mining effort \cite{Valdivia-Berroeta_2022}} \\
PNPA\cite{Rader:22}& Neutral& 113& (010)&[101]& $Cc$ & $d_{333}$ = 591 & $d_{113}$ = 89 \\
NMBA\cite{Ludlow_2024}& Neutral& 23.8& (010)&[101]& $Pc$ & $d_{111}$ = 112& $d_{133}$ = 23\\
ZPAN \cite{ZZZPAN2026}& Neutral& 20.8& (010), (100) & [001]& $Pna2_1$ & $d_{333}$ = 118 & $d_{223}$ = 57 \\
\bottomrule
\multicolumn{8}{c}{Newly Identified Potential NLO Crystals} \\
ASEKUH& Ionic	 & 	208	 & 	(010)	 & 	[101]	 & 	$Ia$	 & 	$d_{333}$	 = 	1096	 & 	$d_{113}$	 = 	159	 \\
LIGXEG& Ionic	 & 	136	 & 	(100),(010)	 & 	[001]	 & 	$Pna2_1$	 & 	$d_{333}$	 = 	769	 & 	$d_{113}$ 	 = 	236	 \\
XAFTEE& Ionic	 & 	106	 & 	(010)	 & 	[001]	 & 	$Cc$	 & 	$d_{333}$	 = 	590	 & 	$d_{113}$	 = 	32	 \\
CUDSAX& Neutral	 & 	91.6	 & 	(010)	 & 	[001]	 & 	$Cc$	 & 	$d_{333}$	 = 	487	 & 	$d_{113}$	 = 	56	 \\
XEGMEB& Ionic	 & 	85.2	 & 	(010)	 & 	[100]	 & 	$Cc$	 & 	$d_{111}$	 = 	362	 & 	$d_{133}$	 = 	280	 \\
HUYRIC& Ionic	 & 	77.9	 & 	(010)	 & 	[100]	 & 	$Cc$	 & 	$d_{111}$	 = 	330	 & 	$d_{133}$	 = 	257	 \\
GUSDEG& Ionic	 & 	69.2	 & 	(001)	 & 	[110]	 & 	$P1$	 & 	$d_{111}$	 = 	336	 & 	$d_{122}$	 = 	116	 \\
\end{tabular}
\footnotetext[1]{Reported values at 1542 nm from Ref \cite{Meier_1998}}
\footnotetext[2]{Reported values at 1900 nm from Ref \cite{Hunziker:08}}
\end{ruledtabular}
\end{table*}

In addition, we verified the accuracy of the calculated $d_{IJK}$ tensors by examining their tensor structures. The structures of the $d_{IJK}$ tensors---specifically the distribution of nonzero elements and equivalent entries---are governed by the crystallographic space group symmetry. Even within the same crystal system, the symmetry constraints dictate that specific $d_{IJK}$ elements may be zero for one space group while remaining significant for another. Therefore, for each crystallographic space group, we compared the calculated $d_{IJK}$ tensors with the symmetry-permitted forms derived from theoretical analysis based on crystallographic space groups. For example, the NLO crystal DAST is crystallized in the space group Monoclinic Cc, its symmetry-allowed $d_{IJK}$ tensor structure is:
\begin{center}
$\begin{bmatrix}
  d_{11} & d_{12} & d_{13} & 0 & d_{15} & 0 \\
  0 & 0 & 0 & d_{24} & 0 & d_{26} \\
  d_{31} & d_{32} & d_{33} & 0 & d_{35} & 0 \\
\end{bmatrix}$
\end{center}

Compared to the $d_{IJK}$ tensor (in the unit of pm/V) obtained through our calculation:
\begin{center}
$\begin{pmatrix}
-309.1 & -47.14 & 4.60 & 0 & 23.48 & 0 \\
0 & 0 & 0 & 5.06 & 0 & -47.14 \\
23.48 & 5.06 & -2.29 & 0 & 4.60 & 0
\end{pmatrix}$
\end{center}
\begin{comment}
\begin{center}
$\begin{pmatrix}
-207.23 & -31.61 & 3.09 & 0 & 15.74 & 0 \\
0 & 0 & 0 & 3.39 & 0 & -31.61 \\
15.74 & 3.39 & -1.54 & 0 & 3.09 & 0
\end{pmatrix}$
\end{center}
\end{comment}

The structures of these tensors are identical, where the positions of the zero and nonzero elements are the same. Moreover, we also observed the relation between the tensor elements, including $d_{12}$ = $d_{26}$, $d_{13}$ = $d_{35}$, $d_{15}$ = $d_{31}$ and $d_{24} = d_{32}$, indicating that the calculated $d_{IJK}$ tensor satisfies the Kleinman symmetry. This validation was performed for all crystallographic space groups, demonstrating consistency between the calculated tensor values and theoretical predictions across all space groups (See section \ref{Checking the structures of Nonlinear Coefficients} of the supplementary information for analysis detail).

Furthermore, we verified our computation results with experimentally measured THz generation efficiencies. The electric field strength of the generated THz radiation ($E_{THz}$) is directly related to the second derivative of induced electric polarization ($P^{NL}$) with respect to time \cite{Murgan:02, Hoffmann_2011}.
\begin{equation}
E_{THz}(t) \propto \frac{d^2P^{NL}(t)}{dt^2}
\label{eq:2detection}
\end{equation}

This fundamental relationship allows us to validate the calculated $P^{NL}$ values by comparing the $P^{NL}$ values with the experimentally measured THz electric field. As shown in Eq. \ref{Ch3_5P_vector}, the $P^{NL}$ values are also crystal-face-dependent, as crystal face dictates the direction of the interaction between the driving light source and the molecular hyperpolarizabilities. Therefore, we calculated the $P^{NL}$ for the three main crystal faces, and reported the maximum $P^{NL}$ value as $P_{max}$ in table \ref{Ch3:Cal_d}, along with the corresponding crystal face.

The predicted crystal faces for optimal THz generation matched the experimental main growth direction for all crystals listed in table \ref{Ch3:Cal_d} \cite{Brunner:08, Li_2014, RAO2016777, Rader_2022, Ho_2022, Ludlow_2024, ZZZPAN2026}. Figure \ref{THz Efficiency of DAST, OH1 and PNPA} compares the trend of the calculated $(P_{max})^2$ and the measured efficiency of known THz generators. The THz conversion efficiency was measured through the intensity of the generated THz pulse, where $I_{THz} \propto E_{THz}^2 \propto (P_{max})^2$. So in Figure \ref{THz Efficiency of DAST, OH1 and PNPA}, we compare the predicted $(P_{max})^2$ to the experimental THz conversion efficiency. The experimental THz conversion efficiency for the generated THz radiation follow the order PNPA $>$ DAST $>$ OH1 (Figure \ref{THz Efficiency of DAST, OH1 and PNPA}), matching the ranking of the calculated $P_{max}$ values. Moreover, the calculated $P_{max}$ values for ZPAN and OH1 are lower than those of DAST and PNPA, consistent with the experimentally observed THz generation efficiencies \cite{Valdivia-Berroeta_2022}.

\begin{figure}[h]
    \centering
    \includegraphics[width=1\linewidth]{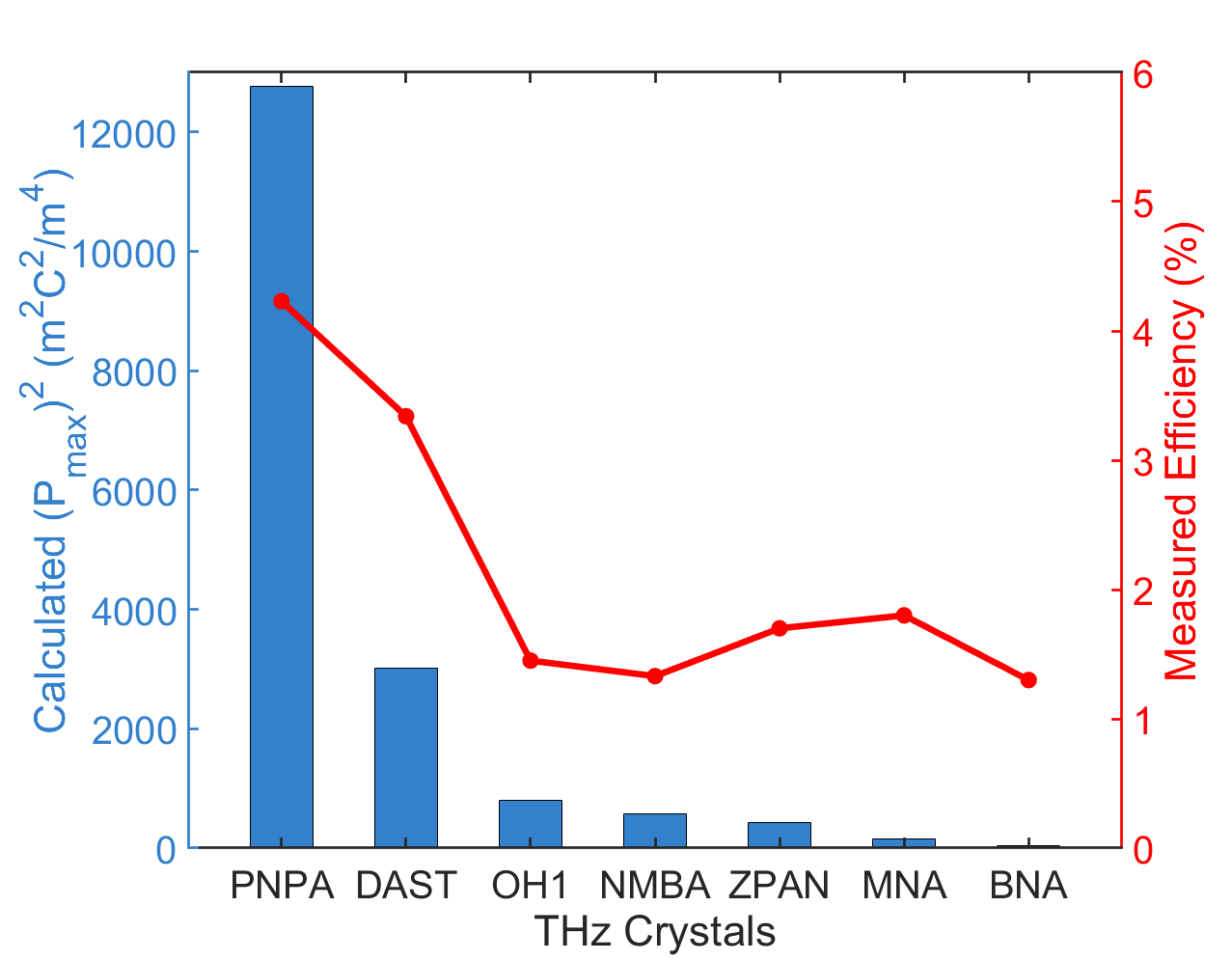}
    \caption{The calculated $(P_{max})^2$ and measured THz Efficiency of known THz generators. The THz efficiency of PNPA, DAST, OH1 and ZPAN were measured at 1450-nm pump wavelength \cite{Rader_2022,ZZZPAN2026}. The THz efficiency of NMBA, MNA and BNA were measured at 1250-nm pump wavelength \cite{Ludlow_2024}.}
    \label{THz Efficiency of DAST, OH1 and PNPA}
\end{figure}
%\hfill

We also notice that the THz generation performances of ZPAN and OH1 vary between studies: OH1 appeared to be more efficient in THz generation Ref \cite{Valdivia-Berroeta_2022}, while ZPAN showed higher efficiency in Ref \cite{ZZZPAN2026}. This difference in THz generation is expected because their $P_{max}$ values are very similar, thus other factors – such as phase matching and the quality of the crystals – are more critical to experimental THz generation outputs. This is supported by the improved THz generation performance of ZPAN after optimization of crystal growth \cite{ZZZPAN2026}, compared to our initial discovery in the data mining work \cite{Valdivia-Berroeta_2022}.

After validating the performance predictions with known THz generators, we are able to identify several potential NLO candidates based on the calculated $P_{max}$ values (Table \ref{Ch3:Cal_d}). The $P_{max}$ values of several newly identified structures are larger than those of the commonly used THz generators, predicting their effectiveness in NLO application. In our previous work \cite{Rader_2022}, we have shown that PNPA is one of the strongest THz generators with a THz generation field strength higher than the state-of-the-art DAST crystal; the calculated $P_{max}$ of PNPA is 113 $mC/m^2$, compared to 54.9 $mC/m^2$ for DAST. The calculated $P_{max}$ for the crystal ASEKUH and LIGXEG are even higher (208 and 136 $mC/m^2$ respectively), making them higher ranked candidates with great potential for high-field THz generation.

%%%%%%%%%%%%%%%%%%%%%%%%%%%%%%%%%%%%%%%%%%%%%%%%%%%%%%%%%%

\subsection{Comparing the accuracy of computed diagonal ($d_{III}$) and off-diagonal ($d_{IJJ}$) nonlinear optical coefficients with other commonly-used simplified approximations}
\textbf{(1) Common approximations to the complete model (Eq. \ref{Ch3eq:d_Full})}

In the previous section, we discussed and validated the NLO values obtained using DFT calculations of the molecular hyperpolarizability and Equations \ref{Ch3eq:d_Full}-\ref{Ch3:chi2_d} to evaluate the second-order nonlinear susceptibility. To the best of our knowledge, this complete model has not been extensively used. Instead simplified approximations are typically utilized to estimate the nonlinear susceptibility \cite{HongHong_2015,Yang_2024, Used_beta_eff_cal_from_cos3, Figi_2008}. One commonly used approximation simplifies the complete model (Equation \ref{Ch3eq:d_Full}) by treating hyperpolarizability as a scalar and considering only a single angle for the projections between the coordinate systems, under-representing the directional dependence of individual tensor elements \cite{Chen_2015, Lee_2018}. After these simplifications, this approximation becomes:
\begin{equation}
\chi^{(2)} = Nf_{local}\beta_{max}\cos^3(\theta_P)
\label{Ch3eq:OP}
\end{equation}

Here $\beta_{max}$ is the magnitude of the molecular hyperpolarizability, where $\beta_{max}$ = $\sqrt{\beta_t^2+\beta_j^2+\beta_k^2}$. $\beta_i$, $\beta_j$ and $\beta_k$ are the combined contribution of the hyperpolarizability tensor elements in the i, j and k directions, where $ \beta_i=\beta_{iii}+\frac{1}{3} \sum_{i\neq k}(\beta_{ikk}+\beta_{kik}+\beta_{kki})$ \cite{Kwon_2019}. $N$ is the number density, which is equal to the number of molecules per unit volume. The quantity $\cos^3(\theta_P)$ is known as the crystal packing order parameter, where $\theta_P$ is defined as the angle between the charge transfer axis of the molecule and the polar axis of the crystal \cite{Kim_2012}. Equation \ref{Ch3eq:OP} attempts to calculate the nonlinear susceptibility along the polar axis, or in other words, approximating the diagonal nonlinear susceptibility elements, where the generated THz light and the input driving light source are polarized in the same direction. This approximation simplifies details of crystal packing and neglects the fact that often off-diagonal susceptibility components can be important. Another simplified equation has been proposed to help account for off-diagonal terms when the NLO process induces a polarization in a different direction than the input electric field. This additional simplification is \cite{Jazbinsek_2019, Chen_2015}:
\begin{equation}
\chi^{(2)} = Nf_{local}\beta_{max}\cos(\theta_P)\sin^2(\theta_P)
\label{Ch3eq:OP_Off_dia}
\end{equation}

Through a straightforward comparison between the simplifications in Equations \ref{Ch3eq:OP}-\ref{Ch3eq:OP_Off_dia} and the complete model in Equations \ref{Ch3eq:d_Full}-\ref{Ch3:chi2_d}, we note that both simplified approximations omit a factor of 2. Consequently, the values obtained from these approximations are actually the $d_{IJK}$ elements, instead of the targeted $\chi^{(2)}$ elements.

\begin{figure}[h]
    \centering
    \includegraphics[width=1\linewidth]{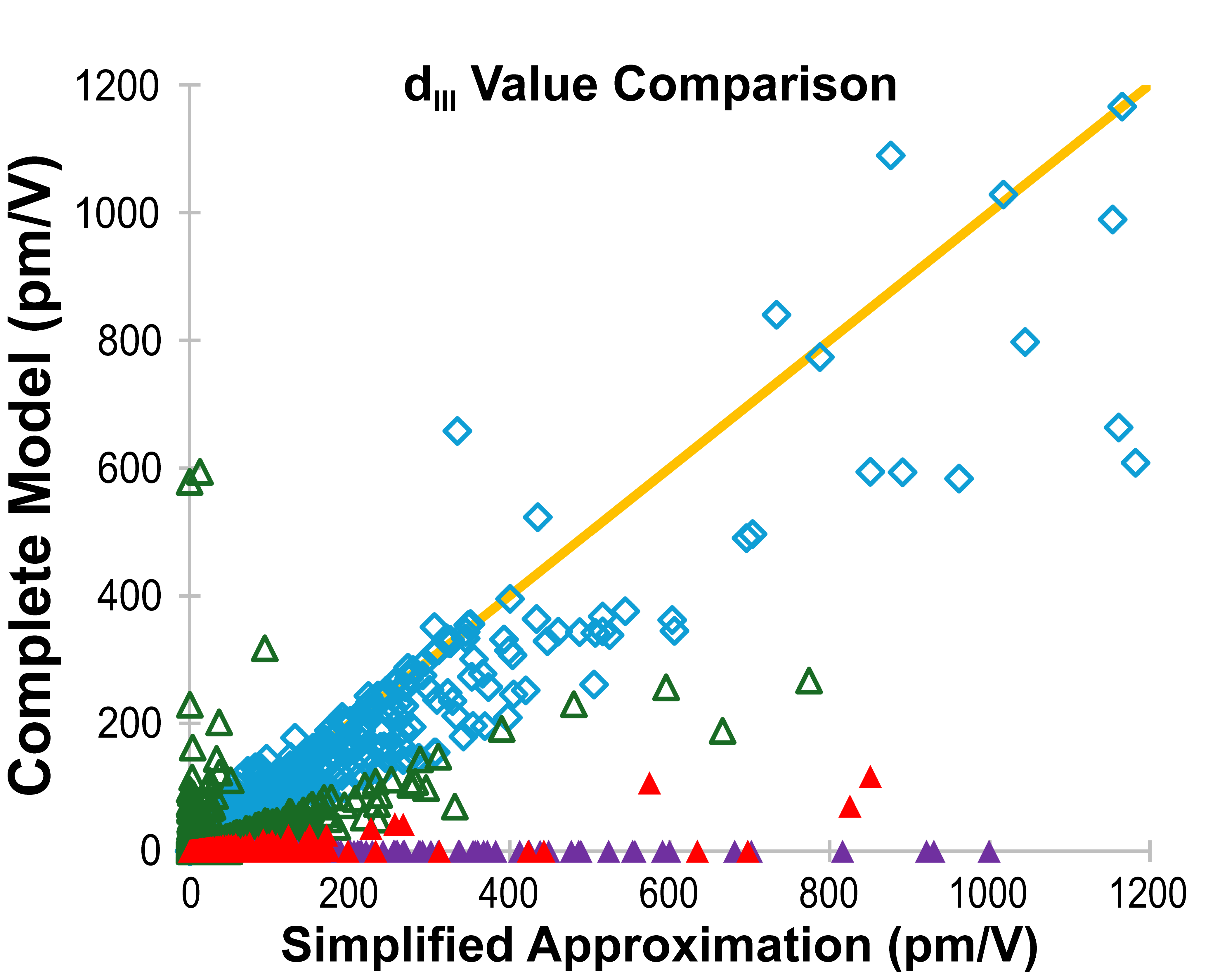}
    \caption{Correlation plot between diagonal element calculated using complete model (Eqs. \ref{Ch3eq:d_Full}-\ref{Ch3:chi2_d}) and simplified approximation (Eq. \ref{Ch3eq:OP}) for all data-mined structures. The yellow line is the identity line, representing the perfect agreement between both models. Blue diamonds indicate a difference within $\pm$50$\%$, while the triangle indicate a difference over $\pm$50$\%$. Within the high-error group, red triangles represent structures with a $\ge$5$\times$ difference, and purple triangles denote a specific low-performance regime where the complete model yields $\le$5pm/V while the simplified approximation overestimates the value at $\ge$50pm/V.}
    \label{Ch3.3:Correlation}
\end{figure}

To examine the accuracy of the most commonly used simplified approximation (Eq. \ref{Ch3eq:OP}), we applied it to the large extracted dataset to compute this alternative diagonal $\chi^{(2)}$ (or the diagonal $d_{III}$ elements) and compared the approximated values to the values obtained from the complete model (Eqs. \ref{Ch3eq:d_Full}-\ref{Ch3:chi2_d}). We specifically focused on the diagonal elements ($d_{III}$) because the emitted THz light is polarized parallel to the driving light polarization in most experimental setups, making the $d_{III}$ elements more relevant in most cases. Figure \ref{Ch3.3:Correlation} shows the correlation plot comparing the $d_{III}$ element calculated from the complete model to the values obtained through simplified approximation (Eq. \ref{Ch3eq:OP}) for each material from our data-mining set.

Contrary to our expectation, the values obtained from the simplified approximation shows a weak correlation with the values calculated by the complete model. We expected the vast majority of the structures would have close or even identical calculated values, clustering near the identity line (yellow line) in figure \ref{Ch3.3:Correlation}. However, a substantial amount of structures fall below the identity line, indicating that the simplified approximation tends to overestimate the $d_{III}$ values. Within the dataset, only 56.4$\%$ of the diagonal $d_{III}$ values calculated from the simplified approximation fall within $\pm$50$\%$ of the values obtained from the complete model (represented by blue diamonds in Figure \ref{Ch3.3:Correlation}). On the other hand, the structures with a difference $\geq$ $\pm$50$\%$ are represented as triangles in Figure \ref{Ch3.3:Correlation}. Specifically, filled red triangles indicate a $\geq5\times$ difference, which applies to 381 structures. Purple triangles denote a distinct group where the complete model (Eqs. \ref{Ch3eq:d_Full}-\ref{Ch3:chi2_d}) yields values $\leq$ 5 pm/V, but the approximation (Eq. \ref{Ch3eq:OP}) overestimates them at $\geq$50 pm/V, which applies to 12$\%$ of the structures. This discrepancy shows that the assumptions made in the simplified model are not always reasonable. To explore the possible reasons for this discrepancy, we carefully examine the calculated $d_{IJK}$ elements for some known THz generation NLO crystals.

\hfill

\textbf{(2) Investigating discrepancies in calculated $d_{IJK}$ for known THz generators}

\begin{table*}
\caption{Diagonal and off-diagonal tensor elements calculated from the complete model (Eq. \ref{Ch3eq:d_Full}) and the simplified approximation (Eq. \ref{Ch3eq:OP}), in units of pm/V), and structural information.}
\label{Ch3_complete_eva}
\begin{ruledtabular}
\begin{tabular}{ c|c c c c c c c }
\multicolumn{2}{c}{} & DAST & OH1 & PNPA & MNA & NMBA & ZPAN \\ \bottomrule
\multicolumn{2}{c}{Space Group} & Cc & $Pna2_1$ & Cc & Cc & Pc &$ Pna2_1$ \\
\multicolumn{2}{c}{Direction of Polar Axis} & xz & z & xz & xz & xz & z \\
\multicolumn{2}{c}{Largest Unit} & 92$^\circ$ & 90$^\circ$ & 98$^\circ$ & 130$^\circ$ & 110$^\circ$ & 90$^\circ$ \\
\multicolumn{2}{c}{Cell Angle} &  &  &  &  &  &  \\
\multicolumn{2}{c}{Angle between polar axis} & 2.1$^\circ$ & 0$^\circ$ & 20.3$^\circ$
 & 18.4$^\circ$ & 40.0$^\circ$ & 0$^\circ$ \\
\multicolumn{2}{c}{and closest cell axis} &  &  &  &  &  &  \\
\multicolumn{2}{c}{OP Angle} & 2.1$^\circ$ & 0$^\circ$ & 20.3$^\circ$
 & 18.4$^\circ$ & 40.0$^\circ$ & 0$^\circ$ \\
 \bottomrule
Complete  &   Diag: & $d_{111}$ = 309 & $d_{333}$ = 160 & $d_{333}$ = 591  & $d_{111}$ = 66   & $d_{111}$ = 112   & $d_{333}$ = 118 \\
 Model &   Off-Diag: & $d_{122}$ = 47 & $d_{223}$ = 18 & $d_{113}$ = 89  & $d_{133}$ = 7 & $d_{133}$ = 23  & $d_{223}$ = 57 \\
\hline
Simplified   & Diag: & $d_{111}$ = 306 & $d_{333}$ = 155 & $d_{333}$ = 849 & $d_{111}$ = 72 & $d_{111}$ = 251 & $d_{333}$ = 121 \\
   Approx. & Off-Diag: & $d_{122}$ = 44 & $d_{223}$ = 45 & $d_{113}$ = 0.11 & $d_{133}$ = 7 & $d_{133}$ = 29 & $d_{223}$ = 59 \\
\end{tabular}
\end{ruledtabular}
\end{table*}

Table \ref{Ch3_complete_eva} shows both diagonal and off-diagonal d tensor components of known THz generators calculated from the complete model (Eqs. \ref{Ch3eq:d_Full}-\ref{Ch3:chi2_d}) and both simplified approximations (Eqs. \ref{Ch3eq:OP}-\ref{Ch3eq:OP_Off_dia}). For DAST, MNA and ZPAN, the simplified approximation yields accurate results; both diagonal and off-diagonal d tensor elements are in excellent agreement with the values calculated from the complete model, with all values falling within a maximum deviation of 6\%. However, substantial discrepancies between both simplified approximations and the complete model are observed in the other calculations, particularly for PNPA. The calculated $d_{333}$ values for PNPA are over 40$\%$ different (591 pm/V from the complete model compared to 849 pm/V from the simplified approximation for diagonal element (Eq. \ref{Ch3eq:OP})) and the difference in the calculated $d_{133}$ values is even more significant (89 pm/V from the complete model compared to 0.11 pm/V from the simplified approximation for the off-diagonal element (Eq. \ref{Ch3eq:OP_Off_dia})). For OH1 and NMBA, one of the d tensor elements from the simplified approximations gives decent results, but the other d tensor elements (the off-diagonal element $d_{223}$ for OH1 and the diagonal element $d_{111}$ for NMBA) show over 200$\%$ discrepancy from the complete model.

One possible reason for the discrepancies is that the unit cell axes for triclinic, monoclinic, trigonal, and hexagonal systems are not all orthogonal, making the calculated tensor elements---except the elements along the x-axis---not exactly along the main unit cell axes. This can be shown by evaluating the correlation between the accuracy of the calculated value and the deviation of the largest unit cell angle from 90$^\circ$. For instance, all of the unit cell angles of ZPAN and OH1 are 90$^\circ$, and the diagonal component from the simplified approximation gives accurate result. Similarly, DAST has a largest unit cell angle of 92$^\circ$, which is still very close to an orthogonal setting, and the simplified approximation still gives an accurate value. In contrast, PNPA and NMBA have larger angles (98$^\circ$ and 110$^\circ$ respectively), and the simplified approximation (Eq. \ref{Ch3eq:OP}) fails to yield an accurate value for the diagonal element.

Another possible reason for the inaccuracy is that the crystal polar axis of crystal structures in triclinic and monoclinic systems is not along one of the crystallographic axes due to the space group symmetry. In these cases, the calculation accuracy is affected by the angle difference between the polar axis and the crystallographic axis. This can be observed by comparing the calculated nonlinear tensor elements of DAST and PNPA. Even though the polar axes of both crystals are on the xz-plane and both crystals are in the same space group, the polar axis of DAST is only 2.1$^\circ$  from the main crystal axis, whereas the angle for PNPA is much larger (20.3$^\circ$). Thus, the calculated diagonal and off-diagonal elements from both simplified approximations are more accurate for DAST than for PNPA. 

To make matters worse, the above two reasons do not apply to orthorhombic, tetragonal and cubic crystals as they are in orthogonal setting and the polar axis is parallel to one of the crystallographic axes due to its crystal symmetry. Yet a large portion of the diagonal nonlinear elements calculated using the simplified approximation (Eq. \ref{Ch3eq:OP}) are inconsistent with those from the complete model (Eq. \ref{Ch3eq:d_Full}), particularly 63.3$\%$ of orthorhombic crystals with symmetry-allowed diagonal elements (Table \ref{Ch3_Statistic_eva}). This indicates that there are more factors contributing to the discrepancy besides the directionality. To explore this discrepancy more thoroughly, we evaluate both calculations mathematically.

\hfill

\textbf{(3) Mathematical Evaluation on the Discrepancy for the diagonal element}

As we already discuss the impact of the polar axis not along crystallographic direction, we next focus on the diagonal nonlinear element of orthogonal crystal systems for this analysis, namely orthorhombic, tetragonal and cubic crystals. For simplicity, we align the input $\beta_{ijk}$ tensor components with the crystallographic axes ($i \parallel I$, $j \parallel J$ and $k \parallel K$) in the below discussion.

In the complete model shown in Eq. \ref{Ch3eq:d_Full}, the cosine terms account for the angle projection onto the three principal axes. For the diagonal $d_{III}$ component, only the co-directional $\beta_{iii}$ term remains. All other $\beta_{ijk}$ terms vanish because they involve the projection onto the perpendicular directions, yielding to a cosine value of zero. Consequently, the expression for the diagonal $d_{III}$ reduces to:
\begin{equation}
    d_{III} = \frac{n}{V}f_{local}\beta_{iii}
    \label{Ch3eq:d}
\end{equation}

In order to compare the simplified approximation (Eq. \ref{Ch3eq:OP}) to the diagonal term produced by the complete model (Eq. \ref{Ch3eq:d_Full}), we can apply the definition of $\beta_{max}$ and expand the expression:
\begin{equation}
    d_{III} = \frac{n}{V}f_{local}\beta_{max}\cos^3(\theta_P)
    \label{Ch3eq:d2}
\end{equation}
\begin{equation}
    d_{III} = \frac{n}{V}f_{local}\cos^3(\theta_P)\sqrt{\sum_{i,j,k}(\beta_{iii}+\beta_{ijj}+\beta_{ikk})^2}
    \label{Ch3eq:d3}
\end{equation}
\begin{multline}
d_{III} = \frac{n}{V} f_{local} \cos^3(\theta_P) \\
\times\sqrt{\begin{aligned}
&\beta_{iii}^2 + 
(\beta_{ijj} + \beta_{ikk})(2\beta_{iii} + \beta_{ijj} + \beta_{ikk}) \\ 
& + (\beta_{jjj} + \beta_{jii} + \beta_{jkk})^2 + (\beta_{kkk} + \beta_{kjj} + \beta_{kii})^2
\end{aligned}}
\label{Ch3eq:d4}
\end{multline}

Eq. \ref{Ch3eq:d} shows that only the diagonal hyperpolarizability tensor element ($\beta_{iii}$) has impact on the $d_{III}$ term in the complete model. In contrast, Eq. \ref{Ch3eq:d4} shows that the crystal order parameter angle ($\theta_{P}$) and all $\beta_{ijk}$ tensor elements are contributing to the $d_{III}$ value, compared to only the effective diagonal term in the complete model. If the effective diagonal hyperpolarizability tensor element ($\beta_{iii}$ in the case shown above) are relatively larger in magnitude than the rest of the tensor elements, the simplified approximation could result in a relatively accurate value. Hence the simplified model is only equal to the complete model when the crystal order parameter angle and all the nonlinear tensor elements except the effective diagonal term are zero.

However, from the equations above, a straightforward trend of rather being overestimating and underestimating the values is hard to conclude, as in theory, the diagonal component is suppressed as the crystal order parameter angle is large. Therefore, we further perform a statistical analysis on the diagonal elements produced by the simplified approximation (Eq. \ref{Ch3eq:OP}).

\begin{table*}
\caption{Statistical assessment of the accuracy of the simplified approximation across non-centrosymmetric triclinic, monoclinic and orthorhombic point groups. The data are categorized into low-error (Less than 50$\%$) and high-error (Over 50$\%$) regimes. The count and the percentage of the numbers of total crystal structures of each category is shown. For each category, the specific breakdown of subsets ($\theta_P<45^\circ$ and $d_{III}$ being the largest tensor element) are also shown. The key statistical differences of each point groups are in bold.}
\label{Ch3_Statistic_eva}
\begin{tabular}{|c|c|c|c|c|c|c|c|c|}
\hline
 & \multicolumn{4}{c|}{Less than 50$\%$ error} & \multicolumn{4}{c|}{Over 50$\%$ error} \\
\hline
 & \multicolumn{2}{c|}{Total} & ~~$\theta_{OP}<45^\circ$~~ & $d_{III}$ being Largest & \multicolumn{2}{c|}{Total} & ~~$\theta_{OP}<45^\circ$~~ & $d_{III}$ being Largest\\
\hline
 & ~Count~ & ~Percent~ & Percent & Percent & ~Count~ & ~Percent~ & Percent & Percent \\
 \hline
~Triclinic~ & 1480 & 47.4$\%$ & 99$\%$ & \textbf{89.1\%} & 1673 & 52.6$\%$ & 99$\%$ & \textbf{17.9\%}  \\
(Group 1) &  &  &  &  &  &  &  &  \\
\hline
~Monoclinic~& 2402 & 54.8$\%$ & 78.2$\%$ & \textbf{78.0\%} & 1981 & 45.2$\%$ & 56.1$\%$ & \textbf{8.0\%}  \\
(Group m) &  &  &  &  &  &  &  &  \\
\hline
~Monoclinic~ & 8866 & 35.5$\%$ & \textbf{50.2\%} & 59.4$\%$ & 16083 & 64.5$\%$ & \textbf{10.9\%} & 38.9$\%$ \\
(Group 2) &  &  &  &  &  &  &  &  \\
\hline
~~~Orthorhombic~~~& 3275 & 36.7$\%$ & \textbf{52.7\%} & 60.0$\%$ & 5656 & 63.3$\%$ & \textbf{18.0\%} & 42.1$\%$  \\  
(Group mm2) &  &  &  &  &  &  &  &  \\
    \hline
    \end{tabular}
\end{table*}

\hfill

\textbf{(4) Statistical analysis on the discrepancy}

Table \ref{Ch3_Statistic_eva} shows the statistical assessment of the accuracy for the diagonal nonlinear tensor elements calculated by the simplified approximation (Eq. \ref{Ch3eq:OP}). We specifically analyzed triclinic, monoclinic and orthorhombic crystals because they consist of over 95$\%$ of the extracted non-centrosymmetric crystals in the CSD. In addition, we also show the statistics based on all non-centrosymmetric point groups within these crystal systems. We categorized the structures based on the point groups because the symmetry operations across point groups are quite different, thus we want to investigate if the point group symmetry also have an impact on the accuracy. Note that the orthorhombic group 222 was excluded because the diagonal tensor elements of all crystals in this group are zero due to its symmetry (detail of the zero and nonzero elements of crystals in different point groups are illustrated in section \ref{Checking the structures of Nonlinear Coefficients} of the supplementary information). 

The accuracy for the triclinic group 1 and monoclinic group m are very similar, which are both close to $50\%$. For monoclinic group 2 and orthorhombic group mm2, the accuracy decreases to roughly 36$\%$. This again confirms the limitation of the simplified approach. 

In addition, we further divided structures with both high- and low-error into two subcategories, namely $\theta_P<45^\circ$ and $d_{III}$ being the largest tensor element (Table \ref{Ch3_Statistic_eva}): 

For the triclinic group 1, the order parameter angle ($\theta_{OP}$) does not affect the accuracy as both percentages in the high- and low-error are identical. It is because the $\theta_{OP}$ is zero for nearly all triclinic non-centrosymmetric structures, since the symmetry group contain only the identity operation; however, the percentage of structure with $\theta_{OP}<45^\circ$ is less than 100$\%$ in both high- and low-error categories because some asymmetric units contain multiple copies of the same molecule in different orientations. In contrast to $\theta_{OP}$, the relative magnitude of diagonal $d_{III}$ elements compared to off-diagonal elements plays a significant role to the accuracy. For 89.1$\%$ of low-error structures and only 17.9$\%$ of high-error structures, the diagonal $d_{III}$ is the largest tensor element. This shows that the relative magnitudes of diagonal compared to off-diagonal components serve as a key indicator for the accuracy of the simplified approach.

For the monoclinic group m, the percentages of the two subcategories in the low-error dataset are both higher than the percentages in the high-error dataset, suggesting that both subcategories---$\theta_{OP<45^\circ}$ and diagonal $d_{III}$ elements being the largest---contribute to the accuracy of the approximation. However, the percentage difference of structures featured $\theta_{OP}<45^\circ$ between the high- and low-error group is 22.1$\%$ while a substantial percentage difference is observed for the structures with diagonal $d_{III}$ elements being the largest element; the percentage increases from 8.0$\%$ in the high-error category to 78.0$\%$ in the low-error category as shown in column 5 and 9 in table \ref{Ch3_Statistic_eva}. This clearly indicates that relative  magnitude of the diagonal $d_{III}$ component is the primary differentiating factor to the accuracy of the approximation (Eq. \ref{Ch3eq:OP}). 

Similar to monoclinic group m, both subcategories also have higher percentages in the low-error dataset compared to the high-error dataset for the monoclinic group 2 and orthorhombic group mm2. However, the statistics show that $\theta_{OP}$ is a more important determination factor for the accuracy, as the percentage of structures featured $\theta_{OP}<45^\circ$ decreases significantly between the low-error and high-error datasets; the percentages drop from 50.2$\%$ to 10.9$\%$ for monoclinic group m and from 52.7$\%$ to 18.0$\%$ for orthorhombic group mm2. This indicates that when $\theta_{OP}$ is greater than 45$^\circ$, it substantially lowers the accuracy of the approximation.

Our analyses reveal the limitation of the simplified approximation, where the accuracy are close to or less than 50$\%$ for the above point groups and the two analyzed features ($\theta_{OP}$ and $d_{III}$) have noticeable impact on the accuracy. This shows that the simplified approximation (Eq. \ref{Ch3eq:OP}) does not necessarily yield an optimal representation of the complete model (Eq. \ref{Ch3eq:d_Full}). In fact, the given approximation is likely derived from  specific frameworks---such as the structural dimensionality models presented by Zyss and Oudar (1982) \cite{PhysRevA.26.2028}---which established various formulations for different nonlinear tensor terms depending on the crystal point group symmetry. For example, the formula for the diagonal term for a crystal from the point group mm2 is $\beta_{yyy}cos^3\alpha$, whereas the formula for the same diagonal term is -$\beta_{yyy}sin^3\alpha$ for the point group m ($\alpha$ is the angle from the symmetry axis to the molecular plane). Consequently, a single form cannot be reliably applied for nonlinear tensor terms across all crystal types.

\section{Conclusion}
In conclusion, we extracted non-centrosymmetric crystal structures from over 1.2 million entries from the CSD, resulting in 77,843 structures. To ensure data validity, we cleansed the dataset and isolated individual molecular structures and performed hyperpolarizability DFT calculation for each molecule. Through DFT calculation and mathematical model, we computed the 2nd order nonlinear susceptibility tensor elements ($\chi^{(2)}_{IJK}$) and the induced dielectric polarization ($P^{NL}$) of all remaining 76,175 structures. The calculated nonlinear elements are consistent with the reported values. The calculated dielectric polarization values are also consistent with the THz generation experiment. Based on our computational results, we have identified several crystal structures -- originally designed for non-NLO applications -- that exhibit comparable or even higher induced dielectric polarization values than well-known THz generators. This highlights their strong potential for NLO applications. We also further compared the nonlinear tensor elements obtained through the complete model to the values estimated from the simplified approximation. Even though some values are in decent agreement with the complete model, we observe a major discrepancy between them across the whole dataset. This lack of agreement suggests that the assumptions made in the simplified approximation may overlook critical factors influencing the crystal properties, highlighting the importance of incorporating the full set of interactions and parameters in such predictive frameworks. Therefore, the complete model should be the preferred method to evaluate the nonlinearity of NLO crystals.

%%%%%%%%%%%%%%%%%%%%%%%%%%%%%%%%%%%%%%%%%%%%%%%%%%%%%%%%%%
%%%%%%%%%%%%%%%%%%%%%%%%%%%%%%%%%%%%%%%%%%%%%%%%%%%%%%%%%%
%%%%%%%%%%%%%%%%%%%%%%%%%%%%%%%%%%%%%%%%%%%%%%%%%%%%%%%%%%

%\vspace*{\fill}
%\newpage
\clearpage

\section{Supplementary Information}
\label{Support_Info}
\subsection{Hyperpolarizability DFT Calculation}
The molecular hyperpolarizability tensors of all mined structures are calculated using Gaussian 16 with basis set 6-31+G** for lighter atoms (Period 1-3) and LANL2DZ for heavier atoms (Period 4-5). By default, the program always attempts to find symmetry within the molecule to reduce the calculation, which results in coordinate transformation. However, the directionality of the calculated tensor elements is important generating the symmetry equivalents parts and the projecting them onto the crystallographic frame to model the second order electric susceptibility. To resolve this problem, we disable the symmetry function in the program, which increased the computation time but ensured the calculated hyperpolarizability tensor was in the input coordinate.

\subsection{Nonlinear Tensor calculation}
In Eq. \ref{Ch3eq:d_Full}, the summation term is over all molecules within a unit cell. Instead of computing hyperpolarizability for every single molecule in the unit cell, we reduced the number of computations by transforming the calculated tensor elements of a single molecule to the direction of other molecular parts based on the crystal symmetry. For example, for crystals with mirror symmetry across the xz-plane, the tensor elements along the y-direction for two symmetry-related molecules should be equal in magnitude but opposite in sign: $\beta_{yyy}$(1st molecule) = -$\beta_{yyy}$(2nd molecule). Once, the hyperpolarizability tensor elements of all the molecules inside the unit cell are obtained, the cosine projection terms in Eq. \ref{Ch3eq:d_Full} are performed after the coordinate transformation described below.
The nonlinear susceptibility calculated in Eq. \ref{Ch3eq:d_Full} is a 3x6 tensor, which represents the correlation of nonlinear optical effects from different crystallographic directions. The nonlinear susceptibility tensor is reported in cartesian coordinate, in contrast, the crystal structure and the atomic positions are commonly reported in fractional coordinate. Therefore, systematically mapping the fractional coordinate onto a fixed cartesian coordinate is vital. There are infinite ways to define a cartesian coordinate, we aimed to simplify and maintain uniformity,  therefore we chose a cartesian system where x = a and b is on the xy-plane:

\begin{gather}
 \begin{bmatrix}
 x \\
 y \\
 z \\
 \end{bmatrix}
 =
  \begin{bmatrix}
  1 & cos(\beta) & cos(\beta) \\
  0 & sin(\gamma) & \frac{cos(\alpha)-cos(\beta)cos(\gamma)}{sin(\gamma)} \\
  0 & 0 & \frac{\Omega}{abcsin(\gamma)}\\
   \end{bmatrix}
  \begin{bmatrix}
  a \\
  b \\
  c \\
   \end{bmatrix}
   \label{Ch3eq:Fractional2Cartesian}
\end{gather}
\begin{equation}
\Omega = abc\sqrt{\begin{aligned}
1-cos^2(\alpha)-cos^2(\beta)-cos^2(\gamma) \\
+2cos(\alpha)cos(\beta)cos(\gamma)
\end{aligned}}
\label{Ch3eqf2c}
\end{equation}

Eq. \ref{Ch3eq:Fractional2Cartesian} and \ref{Ch3eqf2c} map the fractional positions of any atoms onto a cartesian space using the crystallographic information. For crystal system with all 90$^\circ$ unit cell angles --  namely Orthorhombic, Tetragonal and Cubic -- the directionality of fractional coordinate and cartesian coordinate are identical. And for other crystal system, namely Triclinic, Monoclinic, Hexagonal and Trigonal, a-axis in fractional coordinate is directionally equivalent to the x-axis in cartesian and b-axis in fractional coordinate is on the xy-plane in cartesian. Thus, we obtain all the values necessary for Eq. \ref{Ch3eq:d_Full}, except the local field factors. The local field factors are related to the refractive indices of each crystallographic directions: $f_I^\omega = \frac{1}{3}[(n_I^\omega)^2+2]$. Since not all the refractive index values have been reported, thus we estimate the refractive index values based on the measured refractive index values for one of the standard THz crystals – DAST \cite{Jazbinsek_2008}. (Besides measurement, the refractive index values can also be obtained through solid-state DFT calculation, however this type of calculation requires at least an order of magnitude higher computation power than molecular calculation, thus it is not feasible to perform this type of calculation on such a large dataset.)

\subsection{Checking the Structures of Nonlinear Coefficients}
\label{Checking the structures of Nonlinear Coefficients}
To verify the calculation result, we first check the structures of the calculated tensors. Even though there are 18 elements in the $\chi^{(2)}_{IJK}$ tensor, not all of them are independent, and some of them could even be zero based on the crystal symmetry. Before taking space group symmetry into account (which is equivalent to triclinic point group 1), the $\chi^{(2)}$ tensor for any crystals can be expressed in contracted notation:

\begin{flalign}
\chi^{(2)}_{IL}
=
\begin{bmatrix}
  \chi^{(2)}_{11} & \chi^{(2)}_{12} & \chi^{(2)}_{13} & \chi^{(2)}_{14} & \chi^{(2)}_{15} & \chi^{(2)}_{16} \\
  \chi^{(2)}_{21} & \chi^{(2)}_{22} & \chi^{(2)}_{23} & \chi^{(2)}_{24} & \chi^{(2)}_{25} & \chi^{(2)}_{26} \\
  \chi^{(2)}_{31} & \chi^{(2)}_{32} & \chi^{(2)}_{33} & \chi^{(2)}_{34} & \chi^{(2)}_{35} & \chi^{(2)}_{36} \\
   \end{bmatrix} \nonumber \\
\to\
  \begin{bmatrix}
  \chi^{(2)}_{11} & \chi^{(2)}_{12} & \chi^{(2)}_{13} & \chi^{(2)}_{14} & \chi^{(2)}_{31} & \chi^{(2)}_{21} \\
  \chi^{(2)}_{21} & \chi^{(2)}_{22} & \chi^{(2)}_{23} & \chi^{(2)}_{31} & \chi^{(2)}_{14} & \chi^{(2)}_{12} \\
  \chi^{(2)}_{31} & \chi^{(2)}_{32} & \chi^{(2)}_{33} & \chi^{(2)}_{23} & \chi^{(2)}_{13} & \chi^{(2)}_{14} \nonumber
   \label{Ch3_5P_vector}
   \end{bmatrix}
\end{flalign}

\begin{center}
    $I: 1 \to\ x,\ 2 \to\ y,\ 3 \to\ z,$ \\
    $L:1 \to\ xx,\ 2 \to\ yy,\ 3\to\ zz,$
    $\ 4 \to\ yz,\ 5 \to\ xz,\ 6 \to\ xy$
\end{center}

Under the Kleinman symmetry condition, the 18 $\chi^{(2)}$ tensor elements can be further reduced to 10 independent elements. In addition, based on the crystal symmetry, some of the independent elements could be zero. Therefore, the structures of the calculated $\chi^{(2)}$ tensor could be used as an indicator for the validity of the calculation.

For all space groups from our crystallographic dataset, we compared the structures of the calculated $\chi^{(2)}$ with the theoretical tensor structures, our calculated tensors match the theoretical tensor structures based on the Kleinman symmetry. Since monoclinic and orthorhombic crystals make up close to 90$\%$ of the data, below we show the specific results for both crystal systems:

Monoclinic crystal system consists of 3 point groups (2, m, and 2/m). Among these, only group 2 and group m are non-centrosymmetric. On the other hand, orthorhombic crystals also consist of 3 point groups (222, mm2, and mmm), two of these---group 222 and group mm2---are non-centrosymmetric. The forms of the tensors are indicated below:

\begin{center}
\textbf{Monoclinic}  
\end{center}
\begin{center}
\setcounter{MaxMatrixCols}{17}
\begin{tabular}{c}
\textbf{\underline{Group 2}~~~~~~~~~~~~~~~~\underline{Group m}} \\[0.5em]
$\begin{matrix} 
\cdot & \cdot & \cdot & \blacktriangle & \cdot & \bullet & &&&&& \bullet & \blacklozenge & \blacktriangle & \cdot & \blacksquare & \cdot\\ 
\bullet & \star & \blacklozenge & \cdot & \blacktriangle & \cdot & &&&&& \cdot & \cdot & \cdot & \star & \cdot & \blacklozenge\\ 
\cdot & \cdot & \cdot & \blacklozenge & \cdot & \blacktriangle & &&&&& \blacksquare & \star & \cdot & \cdot & \blacktriangle & \cdot 
\end{matrix}$
\end{tabular}
\end{center}

\begin{center}
\textbf{Orthorhombic}  
\end{center}
\begin{center}
\setcounter{MaxMatrixCols}{19}
\begin{tabular}{c}
\textbf{\underline{Group 222}~~~~~~~~~~~~~~\underline{Group mm2}} \\[0.5em]
$\begin{matrix} 
\cdot & \cdot & \cdot & \blacktriangle & \cdot & \cdot & &&&&&&& \cdot & \cdot & \cdot & \cdot & \blacksquare & \cdot\\ 
\cdot & \cdot & \cdot & \cdot & \blacktriangle & \cdot & &&&&&&& \cdot & \cdot & \cdot & \star & \cdot & \cdot\\ 
\cdot & \cdot & \cdot & \cdot & \cdot & \blacktriangle & &&&&&&& \blacksquare & \star & \blacklozenge & \cdot & \cdot & \cdot 
\end{matrix}$
\end{tabular}
\end{center}

The smaller dots represent the positions of the zero elements, and the shapes (circle, star, diamond, triangle and square) represent the non-zero elements. The equivalent elements with equal tensor values are represented using the same shape.

Note that group m, 2, and mm2 all contain diagonal ($\chi^{(2)}$) and off-diagonal terms ($\chi^{(2)}$), which indicates the nonlinear optical effect can be induced by a single polarized driving light source. However, group 222 does not contain any diagonal terms and only contain the mixed terms ($\chi^{(2)}$), indicating that 2 different polarized light sources or a light source polarized between the crystal axes are required to induce the nonlinear optical effect. Moreover, the induced effect is along the direction of the light propagation, thus a more complicated laser setup is required to utilize the nonlinear optical effect for such crystals. Therefore, we will focus on group m, 2, and mm2 below.

A few known THz generators are in group m, namely DAST, PNPA, MNA and NMBA.  Within the point group, DAST, PNPA and MNA are in space group Cc, and NMBA is in space group Pc. The structure of the calculated tensors is the same within the same point group. Below we specifically showed the calculated tensors (in unit of pm/V) for DAST and NMBA, which are from the same point group but with different space groups:

\begin{center}
\textbf{DAST (Group m, Cc)} 
$\begin{pmatrix}
-309.07 & -47.14 & 4.60 & 0 & 23.48 & 0 \\
0 & 0 & 0 & 5.06 & 0 & -47.14 \\
23.48 & 5.06 & -2.29 & 0 & 4.60 & 0
\end{pmatrix}$
\end{center}

\begin{center}
\textbf{NMBA (Group m, Pc)}  
$\begin{pmatrix}
111.65 & 20.43 & 22.83 & 0 & 55.55 & 0 \\
0 & 0 & 0 & 8.52 & 0 & 20.43\\
55.55 & 8.52 & 7.58 & 0 & 22.83 & 0
\end{pmatrix}$
\end{center}

For group 2, known THz generating benzothiazolium crystals, such as PB5FB-T and PBB-T are in this group. However, the structures for the benzothiazolium crystals were not found in the CSD database, and we could not verify the calculation with a known THz generator. Therefore, we selected one of the structures in group 2 from the data mining.

\begin{center}
\textbf{ABAGAN (Group 2, P$2_1$)}  
$\begin{pmatrix}
0 & 0 & 0 & 0.3 & 0 & -1.24 \\
-1.24 & 0.51 & -0.94 & 0 & 0.3 & 0\\
0 & 0 & 0 & -0.94 & 0 & 0.3
\end{pmatrix}$
\end{center}

Lastly, THz generators---OH1,  ZPAN, and BNA---are in group mm 2. And the calculated tensor for OH1 is shown below.

\begin{center}
\textbf{OH1 (Group mm2, $\mathbf{Pna2_1}$)} 
$\begin{pmatrix}
0 & 0 & 0 & 0 & -22.06 & 0 \\
0 & 0 & 0 & -18.16 & 0 & 0 \\
-22.06 & -18.16 & -160.29 & 0 & 0 & 0
\end{pmatrix}$
\end{center}

The above results show that the structures of the calculated tensor are consistent with the theoretical structures based on the Kleinman symmetry.

\clearpage

\begin{comment}
\begin{acknowledgments}
Somthing\dots
\end{acknowledgments}
    
\end{comment}

\begin{comment}
    
\appendix
\section{Appendixes}

\begin{verbatim}
\appendix
\section{}
\end{verbatim}
will produce an appendix heading that says ``APPENDIX A'' and
\begin{verbatim}
\appendix
\section{Background}
\end{verbatim}
will produce an appendix heading that says ``APPENDIX A: BACKGROUND''
(note that the colon is set automatically).
\end{comment}

\bibliography{apssamp}% Produces the bibliography via BibTeX.

@article{Hoffmann_2011,
doi = {10.1088/0022-3727/44/8/083001},
url = {https://dx.doi.org/10.1088/0022-3727/44/8/083001},
year = {2011},
month = {feb},
publisher = {},
volume = {44},
number = {8},
pages = {083001},
author = {Hoffmann, Matthias C and Fülöp, József András},
title = {Intense ultrashort terahertz pulses: generation and applications},
journal = {Journal of Physics D: Applied Physics}
}

@article{Valdivia-Berroeta_2022,
author = {Valdivia-Berroeta, Gabriel A. and Zaccardi, Zachary B. and Pettit, Sydney K. F. and Ho, (Enoch) Sin-Hang and Palmer, Bruce Wayne and Lutz, Matthew J. and Rader, Claire and Hunter, Brittan P. and Green, Natalie K. and Barlow, Connor and Wayment, Coriantumr Z. and Ludlow, Daisy J. and Petersen, Paige and Smith, Stacey J. and Michaelis, David J. and Johnson, Jeremy A.},
title = {Data Mining for Terahertz Generation Crystals},
journal = {Advanced Materials},
volume = {34},
number = {16},
pages = {2107900},
doi = {https://doi.org/10.1002/adma.202107900},
url = {https://advanced.onlinelibrary.wiley.com/doi/abs/10.1002/adma.202107900},
year = {2022}
}

@article{Li_2020,
  title={Metal–Organic Frameworks in Heterogeneous Catalysis: Recent Progress, New Trends, and Future Perspectives},
  author={Li, Bo and Wen, Hui and Cui, Yubo and Zhou, Wei and Qian, Guodong and Chen, Banglin},
  journal={Chemical Reviews},
  volume={120},
  number={16},
  pages={8468--8535},
  year={2020},
  publisher={American Chemical Society},
  doi={10.1021/acs.chemrev.9b00685}
}

@article{Brown_2023,
  title={Hydrogel Tissue Bioengineered Scaffolds in Bone Repair: A Review},
  author={Brown, L. and Zhao, Y. and Nguyen, T. and Kumar, S.},
  journal={Molecules},
  volume={28},
  number={20},
  pages={7039},
  year={2023},
  publisher={MDPI},
  doi={10.3390/molecules28207039}
}

@article{Ye_2020,
  title={Recent Progress in Solid Electrolytes for Energy Storage Devices},
  author={Ye, Tingting and Li, Luhe and Zhang, Ye},
  journal={Advanced Functional Materials},
  volume={30},
  number={29},
  pages={2000077},
  year={2020},
  publisher={Wiley},
  doi={10.1002/adfm.202000077}
}

@article{Eswaran_2025,
  title={A Comprehensive Review of MXene-Based Emerging Materials for Energy Storage Applications and Future Perspectives},
  author={Eswaran, Surulivel Gokul and Rashad, Mohamed and Kumar, Alagarsamy Santhana Krishna and EL-Mahdy, Ahmed F. M.},
  journal={Chemistry – An Asian Journal},
  volume={20},
  number={4},
  pages={e202401181},
  year={2025},
  publisher={Wiley},
  doi={10.1002/asia.202401181}
}

@article{Franken_1961,
  title={Generation of Optical Harmonics},
  author={Franken, P. A. and Hill, A. E. and Peters, C. W. and Weinreich, G.},
  journal={Physical Review Letters},
  volume={7},
  number={4},
  pages={118--119},
  year={1961},
  publisher={American Physical Society},
  doi={10.1103/PhysRevLett.7.118}
}

@Article{Verbiest_1997,
author ="Verbiest, Thierry and Houbrechts, Stephan and Kauranen, Martti and Clays, Koen and Persoons, André",
title  ="Second-order nonlinear optical materials: recent advances in chromophore design",
journal  ="J. Mater. Chem.",
year  ="1997",
volume  ="7",
issue  ="11",
pages  ="2175-2189",
publisher  ="The Royal Society of Chemistry",
doi  ="10.1039/A703434B",
url  ="http://dx.doi.org/10.1039/A703434B"}

@article{Ivanova_2010,
  title={Noncentrosymmetric Crystals with Marked Nonlinear Optical Properties},
  author={Ivanova, Bojidarka B. and Spiteller, Michael},
  journal={The Journal of Physical Chemistry A},
  volume={114},
  number={15},
  pages={5099--5103},
  year={2010},
  publisher={American Chemical Society},
  doi={10.1021/jp911821g}
}

@book{Boyd_2008,
  title={Nonlinear Optics},
  author={Boyd, Robert W.},
  edition={3rd},
  year={2008},
  publisher={Academic Press},
  address={Burlington, MA},
  isbn={9780123694706}
}

@article{PhysRevA.26.2028,
  title = {Relations between microscopic and macroscopic lowest-order optical nonlinearities of molecular crystals with one- or two-dimensional units},
  author = {Zyss, J. and Oudar, J. L.},
  journal = {Phys. Rev. A},
  volume = {26},
  issue = {4},
  pages = {2028--2048},
  numpages = {0},
  year = {1982},
  month = {Oct},
  publisher = {American Physical Society},
  doi = {10.1103/PhysRevA.26.2028},
  url = {https://link.aps.org/doi/10.1103/PhysRevA.26.2028}
}

@article{hubner1994,
  author    = {W. Hübner and K. H. Bennemann and K. Böhmer},
  title     = {Theory for the nonlinear optical response of transition metals: Polarization dependence as a fingerprint of the electronic structure at surfaces and interfaces},
  journal   = {Physical Review B},
  volume    = {50},
  number    = {23},
  pages     = {17597--17610},
  year      = {1994},
  doi       = {10.1103/PhysRevB.50.17597},
  url       = {https://journals.aps.org/prb/abstract/10.1103/PhysRevB.50.17597}
}

@incollection{dmitriev1991,
  author    = {V. G. Dmitriev and G. G. Gurzadyan and D. N. Nikogosyan},
  title     = {Nonlinear Optical Properties of Crystals},
  booktitle = {Handbook of Nonlinear Optical Crystals},
  pages     = {1--30},
  publisher = {Springer},
  year      = {1991},
  doi       = {10.1007/978-3-662-13830-4_3},
  url       = {https://link.springer.com/chapter/10.1007/978-3-662-13830-4_3}
}

@article{Jerphagnon_1970,
    author = {Jerphagnon, J. and Kurtz, S. K.},
    title = {Maker Fringes: A Detailed Comparison of Theory and Experiment for Isotropic and Uniaxial Crystals},
    journal = {Journal of Applied Physics},
    volume = {41},
    number = {4},
    pages = {1667-1681},
    year = {1970},
    month = {03},
    issn = {0021-8979},
    doi = {10.1063/1.1659090},
    url = {https://doi.org/10.1063/1.1659090},
}

@article{Vicario:15,
author = {C. Vicario and M. Jazbinsek and A. V. Ovchinnikov and O. V. Chefonov and S. I. Ashitkov and M. B. Agranat and C. P. Hauri},
journal = {Opt. Express},
number = {4},
pages = {4573--4580},
publisher = {Optica Publishing Group},
title = {High efficiency THz generation in DSTMS, DAST and OH1 pumped by Cr:forsterite laser},
volume = {23},
month = {Feb},
year = {2015},
url = {https://opg.optica.org/oe/abstract.cfm?URI=oe-23-4-4573},
doi = {10.1364/OE.23.004573},
}

@article{Lu:15,
author = {Jian Lu and Harold Y. Hwang and Xian Li and Seung-Heon Lee and O-Pil Kwon and Keith A. Nelson},
journal = {Opt. Express},
number = {17},
pages = {22723--22729},
publisher = {Optica Publishing Group},
title = {Tunable multi-cycle THz generation in organic crystal HMQ-TMS},
volume = {23},
month = {Aug},
year = {2015},
url = {https://opg.optica.org/oe/abstract.cfm?URI=oe-23-17-22723},
doi = {10.1364/OE.23.022723},
}

@article{Meier_1998,
    author = {Meier, U. and Bösch, M. and Bosshard, Ch. and Pan, F. and Günter, P.},
    title = {Parametric interactions in the organic salt 4-N,N-dimethylamino-4'-N'- methyl-stilbazolium tosylate at telecommunication wavelengths},
    journal = {Journal of Applied Physics},
    volume = {83},
    number = {7},
    pages = {3486-3489},
    year = {1998},
    month = {04},
    issn = {0021-8979},
    doi = {10.1063/1.366560},
    url = {https://doi.org/10.1063/1.366560},
}

@article{Li_2014,
title = {Crystal growth and terahertz wave generation of organic NLO crystals: OH1},
journal = {Journal of Crystal Growth},
volume = {402},
pages = {53-59},
year = {2014},
issn = {0022-0248},
doi = {https://doi.org/10.1016/j.jcrysgro.2014.04.033},
url = {https://www.sciencedirect.com/science/article/pii/S0022024814003340},
author = {Yin Li and Zhongan Wu and Xinyuan Zhang and Li Wang and Jianxiu Zhang and Yicheng Wu}
}

@article{Hunziker:08,
author = {Christoph Hunziker and Seong-Ji Kwon and Harry Figi and Flurin Juvalta and O-Pil Kwon and Mojca Jazbinsek and Peter G\"{u}nter},
journal = {J. Opt. Soc. Am. B},
number = {10},
pages = {1678--1683},
publisher = {Optica Publishing Group},
title = {Configurationally locked, phenolic polyene organic crystal 2-\{3-(4-hydroxystyryl)-5,5-dimethylcyclohex-2-enylidene\}malononitrile: linear and nonlinear optical properties},
volume = {25},
month = {Oct},
year = {2008},
url = {https://opg.optica.org/josab/abstract.cfm?URI=josab-25-10-1678},
doi = {10.1364/JOSAB.25.001678},
}

@article{Murgan:02,
author = {Rajan Murgan and David R. Tilley and Yoshihiro Ishibashi and Jeff F. Webb and Junaidah Osman},
journal = {J. Opt. Soc. Am. B},
number = {9},
pages = {2007--2021},
publisher = {Optica Publishing Group},
title = {Calculation of nonlinear-susceptibility tensor components in ferroelectrics: cubic, tetragonal, and rhombohedral symmetries},
volume = {19},
month = {Sep},
year = {2002},
url = {https://opg.optica.org/josab/abstract.cfm?URI=josab-19-9-2007},
doi = {10.1364/JOSAB.19.002007},
}

@article{RAO2016777,
title = {{DAST crystal based Terahertz generation and recording of time resolved photoacoustic spectra of N$_2$O gas at 0.5 and 1.5THz bands}},
journal = {Current Applied Physics},
volume = {16},
number = {7},
pages = {777-783},
year = {2016},
issn = {1567-1739},
doi = {https://doi.org/10.1016/j.cap.2016.04.009},
url = {https://www.sciencedirect.com/science/article/pii/S1567173916300840},
author = {K.S. Rao and A.K. Chaudhary and M. Venkatesh and K. Thirupugalmani and S. Brahadeeswaran}
}

@article{Rader_2022,
author = {Rader, Claire and Zaccardi, Zachary B. and Ho, Sin-Hang Enoch and Harrell, Kylie G. and Petersen, Paige K. and Nielson, Megan F. and Stephan, Harrison and Green, Natalie K. and Ludlow, Daisy J. H. and Lutz, Matthew J. and Smith, Stacey J. and Michaelis, David J. and Johnson, Jeremy A.},
title = {A New Standard in High-Field Terahertz Generation: the Organic Nonlinear Optical Crystal PNPA},
journal = {ACS Photonics},
volume = {9},
number = {11},
pages = {3720-3726},
year = {2022},
doi = {10.1021/acsphotonics.2c01336}
}

@article{Ho_2022,
author = {Palmer, Bruce Wayne H. and Rader, Claire and Ho, Enoch Sin-Hang and Zaccardi, Zachary B. and Ludlow, Daisy J. and Green, Natalie K. and Lutz, Matthew J. and Alejandro, Aldair and Nielson, Megan F. and Valdivia-Berroeta, Gabriel A. and Chartrand, Caitlin C. and Holland, Kayla M. and Smith, Stacey J. and Johnson, Jeremy A. and Michaelis, David J.},
title = {Large Crystal Growth and THz Generation Properties of 2-Amino-5-Nitrotoluene (MNA)},
journal = {ACS Applied Electronic Materials},
volume = {4},
number = {9},
pages = {4316-4321},
year = {2022},
doi = {10.1021/acsaelm.2c00592}
}

@article{Ludlow_2024,
author = {Ludlow, Daisy J.H. and Palmer, Bruce Wayne H. and Green, Natalie K. and Ho, Sin-Hang Enoch and Wayment, Coriantumr Z. and Kelleher, Brenan M. and Barlow, Connor D. and Rollans, Olivia N. and Hunter, Brittan P. and Rader, Claire and Lutz, Matthew J. and Manwaring, Tanner and Smith, Stacey J. and Michaelis, David J. and Johnson, Jeremy A.},
title = {Intense THz Generation with New Organic NLO Crystal NMBA},
journal = {Advanced Optical Materials},
volume = {12},
number = {11},
pages = {2302402},
doi = {https://doi.org/10.1002/adom.202302402},
url = {https://advanced.onlinelibrary.wiley.com/doi/abs/10.1002/adom.202302402},
year = {2024}
}

@Article{Jazbinsek_2019,
AUTHOR = {Jazbinsek, Mojca and Puc, Uros and Abina, Andreja and Zidansek, Aleksander},
TITLE = {Organic Crystals for THz Photonics},
JOURNAL = {Applied Sciences},
VOLUME = {9},
YEAR = {2019},
NUMBER = {5},
ARTICLE-NUMBER = {882},
URL = {https://www.mdpi.com/2076-3417/9/5/882},
ISSN = {2076-3417},
DOI = {10.3390/app9050882}
}

@article{Kim_2012,
author = {Kim, Pil-Joo and Jeong, Jae-Hyeok and Jazbinsek, Mojca and Choi, Soo-Bong and Baek, In-Hyung and Kim, Jong-Taek and Rotermund, Fabian and Yun, Hoseop and Lee, Yoon Sup and Günter, Peter and Kwon, O-Pil},
title = {Highly Efficient Organic THz Generator Pumped at Near-Infrared: Quinolinium Single Crystals},
journal = {Advanced Functional Materials},
volume = {22},
number = {1},
pages = {200-209},
doi = {https://doi.org/10.1002/adfm.201101458},
url = {https://advanced.onlinelibrary.wiley.com/doi/abs/10.1002/adfm.201101458},
year = {2012}
}

@article{Lee_2018,
author = {Lee, Seung-Chul and Kang, Bong Joo and Lee, Ji-Ah and Lee, Seung-Heon and Jazbinšek, Mojca and Yoon, Woojin and Yun, Hoseop and Rotermund, Fabian and Kwon, O-Pil},
title = {Single Crystals Based on Hydrogen-Bonding Mediated Cation–Anion Assembly with Extremely Large Optical Nonlinearity and Their Application for Intense THz Wave Generation},
journal = {Advanced Optical Materials},
volume = {6},
number = {10},
pages = {1701258},
doi = {https://doi.org/10.1002/adom.201701258},
url = {https://advanced.onlinelibrary.wiley.com/doi/abs/10.1002/adom.201701258},
year = {2018}
}

@article{Kwon_2019,
    author = {Kwon, O-Pil and Jazbinsek, Mojca and Seo, Jung-In and Choi, Eun-Young and Yun, Hoseop and Brunner, Fabian D. J. and Lee, Yoon Sup and Günter, Peter},
    title = {Influence of phenolic hydroxyl groups on second-order optical nonlinearity at an example of 2,4- and 3,4-dihydroxyl hydrazone isomorphic crystals},
    journal = {The Journal of Chemical Physics},
    volume = {130},
    number = {13},
    pages = {134708},
    year = {2009},
    month = {04},
    issn = {0021-9606},
    doi = {10.1063/1.3100478},
    url = {https://doi.org/10.1063/1.3100478},
}

@article{Brunner:08,
author = {Fabian D. J. Brunner and O-Pil Kwon and Seong-Ji Kwon and Mojca Jazbin\v{s}ek and Arno Schneider and Peter G\"{u}nter},
journal = {Opt. Express},
number = {21},
pages = {16496--16508},
publisher = {Optica Publishing Group},
title = {A hydrogen-bonded organic nonlinear optical crystal for high-efficiency terahertz generation and detection},
volume = {16},
month = {Oct},
year = {2008},
url = {https://opg.optica.org/oe/abstract.cfm?URI=oe-16-21-16496},
doi = {10.1364/OE.16.016496},
}

@Article{Yang_2023,
AUTHOR = {Yang, Ying and Zhang, Xinyuan and Hu, Zhanggui and Wu, Yicheng},
TITLE = {Organic Nonlinear Optical Crystals for Highly Efficient Terahertz-Wave Generation},
JOURNAL = {Crystals},
VOLUME = {13},
YEAR = {2023},
NUMBER = {1},
ARTICLE-NUMBER = {144},
URL = {https://www.mdpi.com/2073-4352/13/1/144},
ISSN = {2073-4352},
DOI = {10.3390/cryst13010144}
}

@article{Rader:22,
author = {Claire Rader and Megan F. Nielson and Brittany E. Knighton and Zachary B. Zaccardi and David J. Michaelis and Jeremy A. Johnson},
journal = {Opt. Lett.},
number = {22},
pages = {5985--5988},
publisher = {Optica Publishing Group},
title = {Custom terahertz waveforms using complementary organic nonlinear optical crystals},
volume = {47},
month = {Nov},
year = {2022},
url = {https://opg.optica.org/ol/abstract.cfm?URI=ol-47-22-5985},
doi = {10.1364/OL.474343},
}

@ARTICLE{Jazbinsek_2008,
  author={Jazbinsek, Mojca and Mutter, Lukas and Gunter, Peter},
  journal={IEEE Journal of Selected Topics in Quantum Electronics}, 
  title={Photonic Applications With the Organic Nonlinear Optical Crystal DAST}, 
  year={2008},
  volume={14},
  number={5},
  pages={1298-1311},
  doi={10.1109/JSTQE.2008.921407}}

@article{EnochHo2025,
    author = {Biggs, Megan F. and Petersen-Barlow, Paige K. and Ho, (Enoch) Sin-Hang and Chartrand, Caitlin C. and Lattin, Abigail J. and Gebhardt, Halle M. and Lutz, Matthew J. and Hom, William J. and Barlow, Connor D. and Phillips, Gus H. and Sorensen, Kailyn M. and Moody, Megan E. and Jentzsch, Brendon and Hong, Yu-Cheng and Rollans, Olivia N. and Richards, Brigham and Jones, Elisha and Roma, Ashton and Shull, Meredith E. and Michaelis, David J. and Johnson, Jeremy A.},
    title = {Terahertz generation of BNA derivatives},
    journal = {Journal of Applied Physics},
    volume = {138},
    number = {19},
    pages = {193102},
    year = {2025},
    month = {11},
    issn = {0021-8979},
    doi = {10.1063/5.0300701},
    url = {https://doi.org/10.1063/5.0300701},
    
}

@article{XUAN2025141180,
title = {Chalcone-based high nonlinearity coefficient crystals for broadband terahertz output applications},
journal = {Journal of Molecular Structure},
volume = {1326},
pages = {141180},
year = {2025},
issn = {0022-2860},
doi = {https://doi.org/10.1016/j.molstruc.2024.141180},
url = {https://www.sciencedirect.com/science/article/pii/S0022286024036858},
author = {Fanghao Xuan and Shoubo Wang and Dongwei Zhai and Zhaoxin Guo and Jinkang Ma and Kai Xu and Qi Chu and Lifeng Cao and Bing Teng}
}

@article{Sekino_2007,
  title = {Polarizability and second hyperpolarizability evaluation of long molecules by the density functional theory with long-range correction},
  author = {Sekino, Hideo and Maeda, Yasuyuki and Kamiya, Muneaki and Hirao, Kimihiko},
  journal = {The Journal of Chemical Physics},
  volume = {126},
  number = {1},
  pages = {014107},
  year = {2007},
  doi = {10.1063/1.2428291},
  publisher = {AIP Publishing}
}

@article{Yanai_2004,
  title = {A new hybrid exchange--correlation functional using the Coulomb-attenuating method (CAM-B3LYP)},
  author = {Yanai, Takeshi and Tew, David P and Handy, Nicholas C},
  journal = {Chemical Physics Letters},
  volume = {393},
  number = {1-3},
  pages = {51--57},
  year = {2004},
  doi = {10.1016/j.cplett.2004.06.011},
  publisher = {Elsevier}
}

@article{Peach_2008,
  title = {Excitation energies in density functional theory: An evaluation and a diagnostic test},
  author = {Peach, Michael JG and Benfield, Peter and Helgaker, Trygve and Tozer, David J},
  journal = {The Journal of Chemical Physics},
  volume = {128},
  number = {4},
  pages = {044118},
  year = {2008},
  doi = {10.1063/1.2831900},
  publisher = {AIP Publishing}
}

@article{Chen_2015,
author = {Chen, Honghong and Ma, Qi and Zhou, Yuqiao and Yang, Zhou and Jazbinsek, Mojca and Bian, Yongzhong and Ye, Ning and Wang, Dong and Cao, Hui and He, Wanli},
title = {Engineering of Organic Chromophores with Large Second-Order Optical Nonlinearity and Superior Crystal Growth Ability},
journal = {Crystal Growth \& Design},
volume = {15},
number = {11},
pages = {5560-5567},
year = {2015},
doi = {10.1021/acs.cgd.5b01216},
URL = {       https://doi.org/10.1021/acs.cgd.5b01216
},
}

@article{Notake2019,
  author    = {Notake, T. and Takeda, M. and Okada, S. and others},
  title     = {Characterization of all second-order nonlinear-optical coefficients of organic N-benzyl-2-methyl-4-nitroaniline crystal},
  journal   = {Scientific Reports},
  volume    = {9},
  pages     = {14853},
  year      = {2019},
  doi       = {10.1038/s41598-019-50951-1},
  url       = {https://doi.org/10.1038/s41598-019-50951-1},
  publisher = {Nature Publishing Group}
}

@article{KLEIN_2003,
title = {Absolute non-linear optical coefficients measurements of BBO single crystal and determination of angular acceptance by second harmonic generation},
journal = {Optical Materials},
volume = {22},
number = {2},
pages = {163-169},
year = {2003},
note = {Proceedings of the Scientific Committe of the French Research Group "GDR 1148 CNRS": LASMAT: Research Group on Laser Materials},
issn = {0925-3467},
doi = {https://doi.org/10.1016/S0925-3467(02)00360-9},
url = {https://www.sciencedirect.com/science/article/pii/S0925346702003609},
author = {R.S Klein and G.E Kugel and A Maillard and A Sifi and K Polgár}
}

@article{Shoji_97,
author = {Ichiro Shoji and Takashi Kondo and Ayako Kitamoto and Masayuki Shirane and Ryoichi Ito},
journal = {J. Opt. Soc. Am. B},
number = {9},
pages = {2268--2294},
publisher = {Optica Publishing Group},
title = {Absolute scale of second-order nonlinear-optical coefficients},
volume = {14},
month = {Sep},
year = {1997},
url = {https://opg.optica.org/josab/abstract.cfm?URI=josab-14-9-2268},
doi = {10.1364/JOSAB.14.002268},
}

@misc{ZZZPAN2026,
      title={ZPAN: An Organic Nonlinear Optical Crystal for High Intensity THz Generation}, 
      author={Matthew J. Lutz and Sin-Hang Ho and Zachary B. Zaccardi and Paige Petersen and Elisha Jones and Stacey J. Smith and David J. Michaelis and Jeremy A. Johnson},
      year={2026},
      eprint={2607.13954},
      archivePrefix={arXiv},
      primaryClass={physics.optics},
      url={https://arxiv.org/abs/2607.13954}, 
}

@article{HongHong_2015,
    author = {Xu, Kai and Ma, Jinkang and Wang, Tianhua and Cao, Lifeng and Teng, Bing},
    title = {New Organic Nonlinear
Optical Pyridinium-Based Hydrate
Crystals: Trimethoxy-Induced Noncentrosymmetric Alignment},
    journal = {Crystal Growth \& Design},
    volume = {23},
    number = {12},
    pages = {9041-9051},
    year = {2023},
    month = {11},
    issn = {1528-7483},
    doi = {10.1021/acs.cgd.3c01081},
    url = {https://doi.org/10.1021/acs.cgd.3c01081},
}

@article{Yang_2024,
author = {Yang, Jeong-A and Lee, Chae-Won and Kim, Chaeyoon and Auer, Michael and Yu, In Cheol and OH, Jungkwon and Yoon, Woojin and Yun, Hoseop and Kim, Dongwook and Jazbinsek, Mojca and Rotermund, Fabian and Kwon, O-Pil},
title = {Chiral Cationic Chromophores: A New Class of Efficient Ultrabroadband Organic THz Crystals},
journal = {Advanced Optical Materials},
volume = {12},
number = {22},
pages = {2400343},
doi = {https://doi.org/10.1002/adom.202400343},
url = {https://advanced.onlinelibrary.wiley.com/doi/abs/10.1002/adom.202400343},
year = {2024}
}

@article{Used_beta_eff_cal_from_cos3,
author = {Shin, Myeong-Hoon and Lee, Seung-Heon and Kang, Bong Joo and Jazbinšek, Mojca and Yoon, Woojin and Yun, Hoseop and Rotermund, Fabian and Kwon, O-Pil},
title = {Organic Three-Component Single Crystals with Pseudo-Isomorphic Cocrystallization for Nonlinear Optics and THz Photonics},
journal = {Advanced Functional Materials},
volume = {28},
number = {48},
pages = {1805257},
doi = {https://doi.org/10.1002/adfm.201805257},
url = {https://advanced.onlinelibrary.wiley.com/doi/abs/10.1002/adfm.201805257},
year = {2018}
}

@article{Figi_2008,
author = {Harry Figi and Lukas Mutter and Christoph Hunziker and Mojca Jazbin\v{s}ek and Peter G\"{u}nter and Benjamin J. Coe},
journal = {J. Opt. Soc. Am. B},
number = {11},
pages = {1786--1793},
publisher = {Optica Publishing Group},
title = {Extremely large nonresonant second-order nonlinear optical response in crystals of the stilbazolium salt DAPSH},
volume = {25},
month = {Nov},
year = {2008},
url = {https://opg.optica.org/josab/abstract.cfm?URI=josab-25-11-1786},
doi = {10.1364/JOSAB.25.001786},
}

@article{Goncharov_2012,
    author = {Goncharov, Vladimir A. and Varga, Kalman},
    title = {Real-space, real-time calculation of dynamic hyperpolarizabilities},
    journal = {The Journal of Chemical Physics},
    volume = {137},
    number = {9},
    pages = {094111},
    year = {2012},
    month = {09},
    issn = {0021-9606},
    doi = {10.1063/1.4749793},
    url = {https://doi.org/10.1063/1.4749793},
}

\end{document}